%% file: ClassSLines.tex
\documentclass[11pt]{article}
\pdfoutput=1

\usepackage[]{Packages}
\bibliography{T-duality}

\usepackage{Definitions}

\usepackage{textcomp}

\preprint{}

\newcommand{\OfficialTitle}{
Line operators from M--branes on compact Riemann surfaces
}
\title{\vspace{2cm}
  {\color{Thoughtless}\Huge\textbf{\dosserif\OfficialTitle}}
}

\hypersetup{pdfauthor={Antonio Amariti and Domenico Orlando and Susanne Reffert},pdftitle={\OfficialTitle}}

\author{%
  \begin{minipage}{.8\linewidth}
    \vspace{1cm}
    \begin{center} \dosserif
      {\small 
        \textbf{Antonio~Amariti}\textsuperscript{\ding{95}},
         \textbf{Domenico~Orlando}\textsuperscript{\ding{96}} and 
        \textbf{Susanne~Reffert}\textsuperscript{\ding{96}}}
    \end{center}
    \vspace{1cm}
     \authorBlock{\ding{95}}{ Physics Department, The City College of the \textsc{cuny},\\
160 Convent Avenue, New York, \textsc{ny} 10031, \textsc{usa}}
    \authorBlock{\ding{96}}{Albert Einstein Center for Fundamental Physics\\
      Institute for Theoretical Physics\\
      University of Bern,\\
      Sidlerstrasse 5, \textsc{ch}-3012 Bern, Switzerland}
  \end{minipage}
}

\date{} 

\begin{document}

\setstretch{1.15}

\numberwithin{equation}{section}

\begin{titlepage}

  \newgeometry{top=23.1mm,bottom=46.1mm,left=34.6mm,right=34.6mm}

  \maketitle

  \thispagestyle{empty}

  \vfill\dosserif
  
  \abstract{\normalfont \noindent
In this paper, we determine the charge lattice 
of mutually local Wilson and 't Hooft line operators
for class S theories living on \M5--branes wrapped on compact Riemann surfaces.
The main ingredients of our analysis are the fundamental group of the $N$--cover of the Riemann surface,
and a quantum constraint on the six-dimensional theory.
This latter plays a central role in excluding some of the possible lattices and 
imposing consistency conditions on the charges.
This construction gives a geometric explanation for the 
mutual locality among the lines, fixing their charge lattice and the structure of the four-dimensional gauge group.
 }

\vfill

\end{titlepage}

\restoregeometry

\section{Introduction}

In this work, we classify the possible gauge groups corresponding to a given gauge algebra for four-dimensional $\mathcal{N}=2$ quiver gauge theories  via an M--theory construction, where $N$ \M5--branes are wrapped on a Riemann surface $\Sigma_{g,0}$ of genus $g>2$.  We find that unlike in the case of $\mathcal{N}=4$ \ac{sym} theory~\cite{Amariti:2015dxa}, which descends from the \M5--branes wrapped on a torus, the topological data of the \M5--brane geometry is \emph{not enough} to classify the theories and derive the \ac{dsz} quantization condition. Due to the reduced symmetry of the case at hand, it becomes necessary to \emph{impose a quantum condition} already in the six-dimensional theory which selects the allowed multiple covers of $\Sigma_{g,0}$ (corresponding to the possible gauge groups) and the allowed lines on $\Sigma_{g,0}$ (corresponding to the allowed line operators in the gauge theory).

\bigskip

Symmetries are one of the main tools for characterizing the spectrum 
of \acp{qft}.
Local symmetries are  redundancies of the theory and are associated to 
the choice of a gauge group.
A fact often overlooked is that the gauge group is not completely fixed by the gauge algebra - additional information is required.
In \ac{qcd}, for example, the additional data is given by charged matter fields transforming in the fundamental
representation. They naturally promote the gauge algebra $A_2$ to the 
universal covering group $SU(3)$.
The matter content does however not always fix the structure of the group completely
for a given algebra.
A simple but instructive example is $\mathcal{N}=4$ \ac{sym} theory with gauge algebra $A_{N-1}$.
In this case, a generic $SU(N)/\mathbb{Z}_k$ gauge group is compatible 
with the existence of matter fields in the adjoint representation of
the gauge algebra and additional data must be specified. A solution to
this problem was 
given in~\cite{Aharony:2013hda}, where it was observed that  
the gauge group is \emph{fixed} by the charge lattice 
of \ac{Wline} and \ac{Hline} operators.
Once the maximal charge lattice of mutually local \ac{WH} bound states is specified, the gauge group is uniquely determined.
In the four-dimensional analysis of~\cite{Aharony:2013hda}, the charge lattices are constructed by imposing 
a \ac{dsz} quantization condition on the \ac{WH} lines.
The net result is the following: the possible lattices are generated by two vectors $(k,0)$ and $(i,k')$,
with $k k' = N$ and $i<k$. We refer to this lattice as \(\Gamma_{N;k,i}\).
The corresponding gauge group has been defined 
 $(SU(N)/\mathbb{Z}_k)_i$ in~\cite{Aharony:2013hda}.

An equivalent construction can be engineered in M--theory,
where the \ac{dsz}  quantization condition results from a purely classical constraint, imposed on the geometry.
This can be  understood by considering M--theory compactified on a torus~\cite{Amariti:2015dxa}.
Geometrically, this theory describes the dynamics on $N$ \M5--branes wrapped on 
a genus one Riemann surface\footnote{We refer to a genus \(g\) Riemann surface with \(n\) puncture as $\Sigma_{g,n}$.}  $T^2=\Sigma_{1,0}$.
When compactified on $T^2$, the theory describes a stack of $N$ \D3--branes in $\mathbb{R}^{1,3}$, \emph{i.e.}  $\mathcal{N}=4$
\ac{sym}. The line operators are \M2--lines
 wrapping a geodesic curve on $T^2$. In \tIIB string theory, they become bound states of 
\F{}--strings and \D{}--strings.
These strings are extended transversally to the stack of \D3--branes 
with one endpoint on the stack and the other at infinity.
They are interpreted in the field theory regime as bound states of \ac{WH} lines.
Their charge lattice can be described in the M--theory language by introducing the notion of the \emph{fundamental group}.\footnote{The necessary notions of algebraic topology are collected in Appendix~\ref{sec:details}.}
This is done as follows. Associate the two directions of $\Sigma_{1,0}$
to the two freely homotopic generating closed curves, $a$ and $b$. The \M2--lines wrap these cycles.
The fundamental group is specified by
\begin{equation}
  \pi_1(T^2) = \braket{ a, b}{\comm{a}{b} = e},
\end{equation}
where $\comm{a}{b} \equiv a b a^{-1} b^{-1}$, and $e$ is the identity, \emph{i.e.} a cycle contractible to a point in the geometry.
In this case, $\pi_1(T^2)$ is Abelian and isomorphic to $\mathbb{Z} \times \mathbb{Z}$. 
Now consider an $N$--fold cover of this surface.
This defines a new Riemann surface, with genus $g'=g=1$ that we denote by \(T^2_{N;k,i}\).
Its fundamental group is given by
\begin{equation}
  \pi_1(T^2_{N;k,i}) \braket{a^k, a^i b^{k'}}{\comm{a^k}{a^i b^{k'}}=e},
\end{equation}
with $k k' = N$ and $i<k$. 
We express the homologies of the lines in the multiple cover in terms of the homologies of the lines in the base. This allows us to read off the charges of the \ac{WH} operators.
In this way, we 
obtain the generating vectors of the charge lattice. 
This reproduces the lattice $\Gamma_{N;k;i}$ and, as a consequence, fixes the
gauge group $(SU(N)/Z_k)_i$.
Note that in our formalism, the \ac{dsz} quantization condition is not \emph{imposed} on the charges but follows as a consequence of the M--theory construction.

\bigskip
In this article, we study the \emph{generalization to higher genus Riemann 
surfaces}.
In gauge theory language, this amounts to studying the global properties and the charge 
lattice of class S theories~\cite{Gaiotto:2009we}.\footnote{See also~\cite{Drukker:2009tz,Xie:2013lca,Tachikawa:2013hya,Xie:2013vfa,Bullimore:2013xsa,Tachikawa:2015iba,Coman:2015lna} for further discussion of this topic.}

They originate as six-dimensional $\mathcal{N}=(2,0)$ theories living on the worldvolume of a stack of $N$ \M5--branes  wrapped on an orientable genus $g$ Riemann surface with $n$ punctures $\Sigma_{g,n}$.
The four-dimensional theories are obtained by a partially twisted compactification on $\Sigma_{g,n}$, giving rise to
a four-dimensional $\mathcal{N}=2$ quiver gauge theory.
The quiver can be described as follows: consider a six-dimensional theory living on a three-punctured sphere 
$\Sigma_{0,3}$. The four-dimensional theory in this case is a strongly coupled $\mathcal{N}=2$ theory,
known as $T_N$ theory, with a classical $SU(N)^3$ flavor symmetry. The $T_N$ theories can be used as building blocks for constructing a theory on 
$\Sigma_{g,n}$. This is done by gluing the punctures of the $T_N$ blocks, which correspond to pairs of pants.
The gluing is associated to the gauging of the $SU(N)$ flavor symmetries. 

In the following, we restrict ourselves to the case of compact Riemann surfaces with $n=0$, 
obtained by gluing all the punctures together (\(g\)-fold torus).
As we will see, this procedure requires each puncture to be maximal,
\emph{i.e.} associated to the full non--Abelian global $SU(N)$ in an $N$-cover.
In summary, the quiver can be reconstructed by specifying  a pants decomposition of the surface.
One can decompose the $\Sigma_{g,0}$ surface into $2(g-1)$ pants $\Sigma_{0,3}$.
This corresponds to having $3(g-1)$ gauge groups.
Observe that different pants decompositions are possible: they specify different topologies and different quivers. These quivers are related to the mapping class group  of $\Sigma_{g,0}$, 
which corresponds to the action of the S--duality group.

We study the charge lattices of these class S theories via the fundamental group as described above for the case of $\mathcal{N}=4$ \ac{sym}.
We find, however, that there are some important qualitative differences between the two cases. 

The fundamental group of the \(N\)--cover has \(2 g' = N (g - 1) + 1\) generators. This necessitates some care for the projection of the freely homotopic closed curves onto the charges of the \ac{WH} bound states.

Another difference is due to the central symmetry of a $T_N$ block. Despite the fact that the flavor symmetry of $T_N$ is $SU(N)$, at quantum level, the central symmetry is $\mathbb{Z}_N$ and not $\mathbb{Z}_N^3$~\cite{Gaiotto:2009we,Gadde:2011ik}.
  When gluing the $2(g-1)$ blocks, the subgroup $\mathbb{Z}_N \subset \mathbb{Z}_N^{2(g-1)}$ 
 remains as a global symmetry of the quantum theory~\cite{Aharony:1998qu,Tachikawa:2013hya}.
 This imposes a quantum constraint at the level of the six-dimensional theory and,
 differently from the $\mathcal{N}=4$ \ac{sym} case, the six-dimensional origin of the 
 four-dimensional charge lattice is intrinsically quantum.
This dramatically reduces the growth of the number of allowed lattices,
from exponential to polynomial growth in $N$.

\bigskip
The paper is organized as follows.
In Section~\refstring{sec:general}, we present our general strategy in the derivation of the
lattices for class S theories on compact Riemann surfaces.
In Section~\refstring{sec:genus2}, we study explicitly the simplest non-trivial example, the double cover
of $\Sigma_{2,0}$.
In Section~\refstring{sec:generalizations}, we discuss the generalization to higher genus and higher multiple covers.
In Section~\refstring{sec:conclusions}, we end with conclusions and further directions.
In Appendix~\ref{sec:details}, we have collected some mathematical details on the fundamental group and its relation to the multiple covers.

\section{General strategy}
\label{sec:general}

In this section, we explain our strategy for extracting the charge lattices of 
class S theories obtained from the partially twisted compactification
of the six-dimensional $(2,0)$ theory on compact Riemann surfaces $\Sigma_{g,0}$.
We consider $N$ \M5--branes wrapped on $\Sigma_{g,0}$. This wrapping
defines a new surface $\Sigma_{g',0}$, where the genus $g'$ is related to $g$ by
\begin{equation}
  \label{eq:newgenus}
  g' = N \pqty{g-1} + 1.
\end{equation}

The line operators of the four-dimensional theory are obtained by considering closed 
\M2 lines in the six-dimensional geometry and reducing on the Riemann surface. They become bound states of (\F1,\D1) strings
in \tIIB string theory, \emph{i.e.}  dyonic \ac{WH} lines in the field theoretical language.
In order to describe the charge lattice of the latter, 
we need to study the intersection theory of the \M2--lines.
This is done by considering the fundamental group of  \(\Sigma_{g',0}\).
For a given Riemann surface \(\Sigma_{g,0}\), we define the inequivalent \(N\)--covers 
$\{$\(\Sigma^N_{g,0}\) $\}$ as the set of all the surfaces of type \(\Sigma_{g',0}\) whose fundamental group \(\pi_1(\Sigma_{g',0})\) is a subgroup of \(\pi_1(\Sigma_{g,0})\) (up to conjugation). 
The fundamental group allows us to describe the \(N\)--cover in terms of probe \M2--branes wrapping closed curves on \(\Sigma^N_{g,0}\). 
This subgroup structure implies that the fundamental group of each \(\Sigma^N_{g,0}\) can be written in terms of the cycles \(a_i\) and \(b_i\) of \(\Sigma_{g,0}\). In the case of \(g=1\), this is precisely the construction in~\cite{Amariti:2015dxa}. The lattices \(\Gamma_{N;k,i}\) are the fundamental groups of the covers, which are all subgroups of the integral lattice \(\setZ\times\setZ\), \emph{i.e.} of the fundamental group of the torus.
For \(g>1 \), the combinatorics is much more intricate and the multiple covers are more easily described in terms of the symmetric group \(S_N\) (see Appendix~\ref{sec:multiple-covers}). Exact formulas exist in the mathematical literature. The number of inequivalent covers grows exponentially with \(N\) (for the exact expression, see~\cite{Lubotzky}):
\begin{equation}
  a(N) \sim 2 N (N!)^{2g - 2}.
\end{equation}

In spite of the obvious similarities with the \(\mathcal{N} = 4\) case, the construction for \(\mathcal{N} = 2\) quiver gauge theories presents a fundamental qualitative difference:
\emph{not all multiple covers of a Riemann surface correspond to a quiver gauge theory of class S}.
In order to select the allowed multiple covers, \emph{an additional quantum constraint has to be imposed already at the level of the six-dimensional theory}.
The nature of this constraint can be understood as follows. Consider a compact Riemann surface and decompose it in terms of pairs of pants, surfaces $\Sigma_{0,3}$ with maximal punctures. 
Each pair of pants has a classical $SU(N)^3$ global symmetry, with center $\mathbb{Z}_N^3$.
At the quantum level, only the diagonal central $\mathbb{Z}_N$ symmetry is preserved~\cite{Gaiotto:2009we,Gadde:2011ik}.
We will therefore consider only those multiple covers that preserve this symmetry. 

\begin{figure}
  \centering
  \begin{tikzpicture}
    \node at (-4,4) {\includegraphics[width=4.5cm]{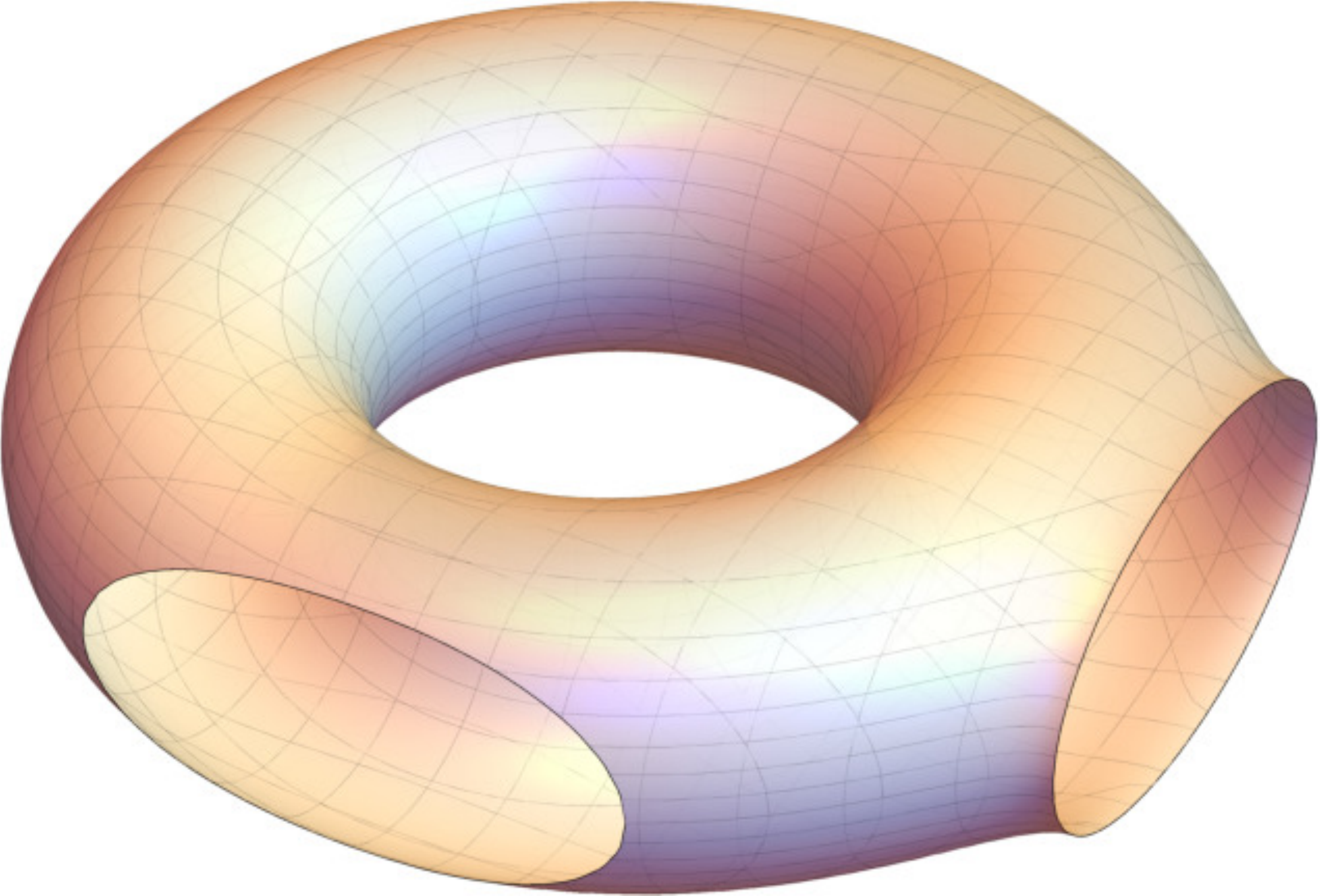}};
    \node at (4,4) {\includegraphics[width=4.5cm]{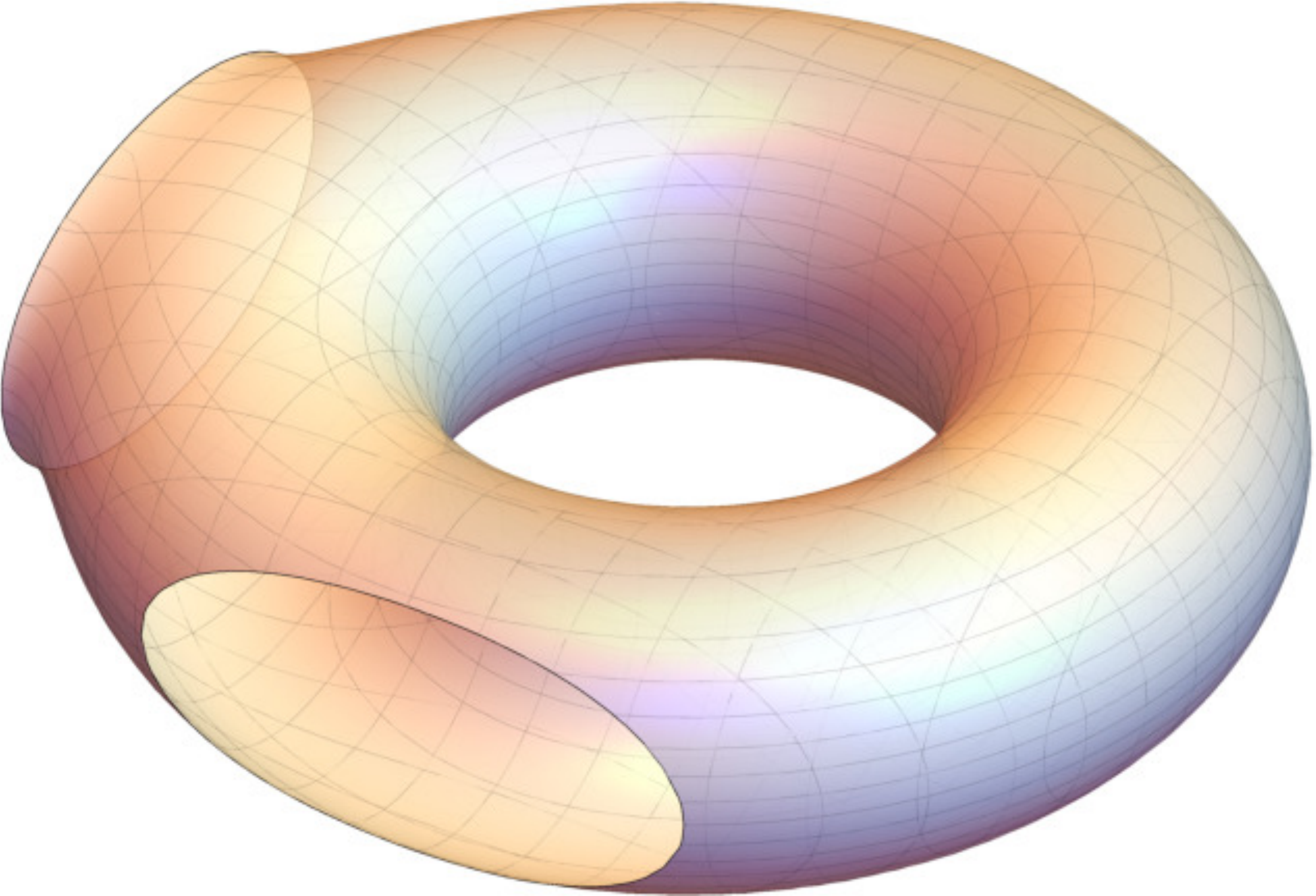}};
    \node at (-4,-.5) {\includegraphics[width=4.5cm]{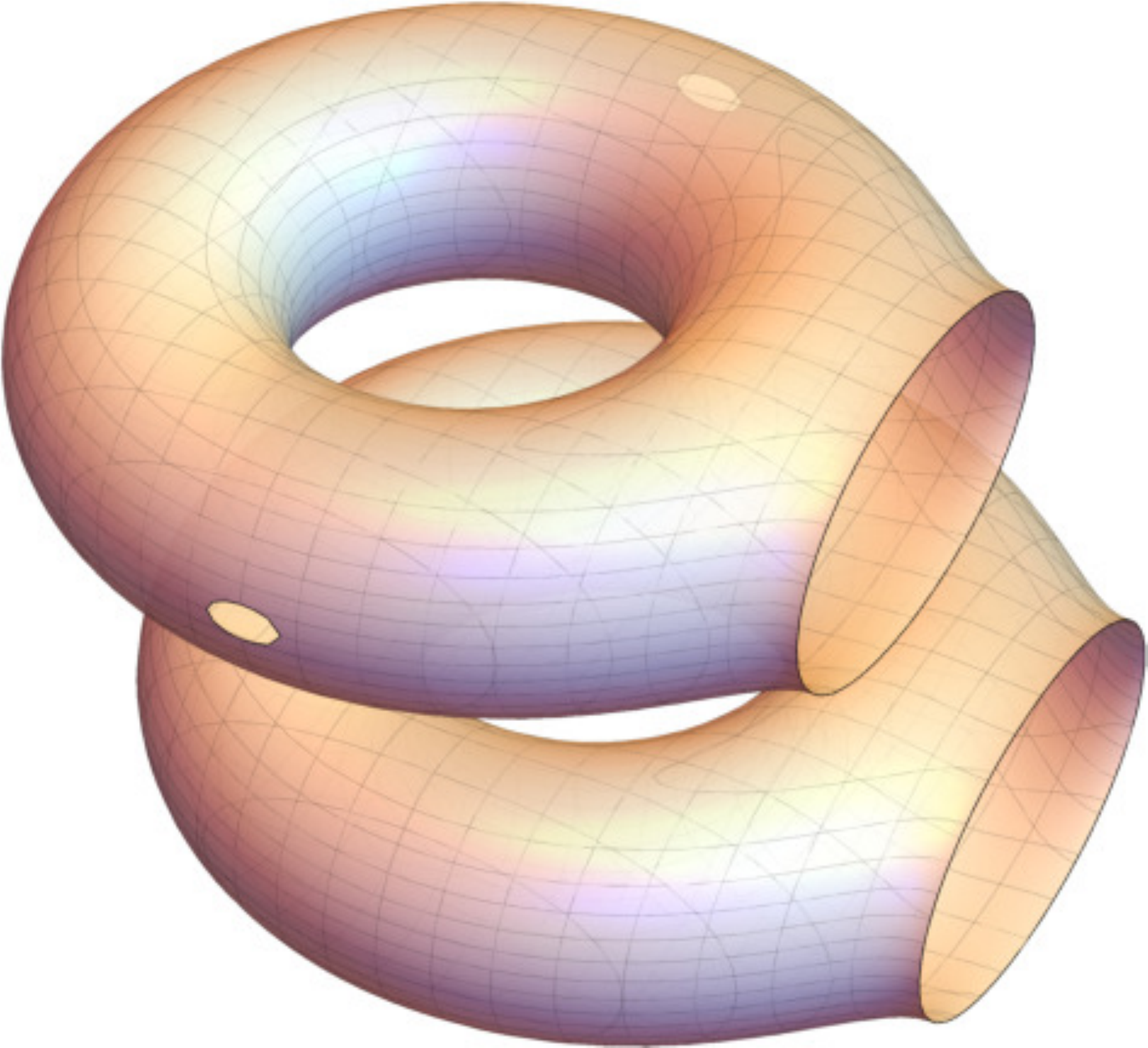}};
    \node at (4,-.5) {\includegraphics[width=4.5cm]{T2-connected-cover-right}};
    \node at (-4,2) {\((\ydiagram{2},\ydiagram{2})\)};
    \node at (4,2) {\((\ydiagram{2},\ydiagram{2})\)};
    \node at (0,2) {(a)};
    \node at (0,-3.5) {(b)};
    \node at (-4,-3) {\((\ydiagram{1,1},\ydiagram{1,1})\)};
    \node at (4,-3) {\((\ydiagram{2},\ydiagram{2})\)};
  \end{tikzpicture}
  \caption{Two types of double covers of the double torus. In (a) we
    have the union of two connected double covers of \(\Sigma_{1,1}\),
    in (b) one of the two covers is disconnected. A cover of type (b) breaks the \(\setZ_N\) symmetry.}
  \label{fig:double-covers-double-torus}
\end{figure}

As discussed in Appendix~\ref{sec:multiple-covers}, we can label the covers by ordered pairs of partitions of $N$, \emph{i.e.} 
one partition per generator of the base.
We distinguish two classes.
In the first class, all the partitions pairs  \((Y(a_i),
Y(b_i))\) are rectangular and the number of columns in one diagram is
greater or equal to the number of rows in the other: these correspond to connected \(N\)--covers of a
torus. In the second class, at
least one of the pairs of partitions $(Y(a_i), Y(b_i))$ describes a
disconnected cover. The two types of covers differ as shown in Figure~\ref{fig:double-covers-double-torus}: let us consider a double cover of $\Sigma_{2,0}$ by cutting it into two genus 1 tori $\Sigma_{1,1}$ and taking their double covers separately before gluing them together to form a genus 3 surface. The first class is made from two tori with two punctures $\Sigma_{1,2}$, \emph{i.e.} two double covers of $\Sigma_{1,1}$; the second class is made from two copies of a torus with one puncture $\Sigma_{1,1}$ (a disjoint double cover of $\Sigma_{1,1}$), and a torus with two punctures $\Sigma_{1,2}$. %
In the presence of maximal punctures, covers of the second class do not respect the quantum constraint because they
break the $Z_N$ symmetry.

The number of allowed covers grows like \(a(N) = (\sigma_1(N))^g\), where \(\sigma_1\) is the divisor function. Asymptotically, this is a polynomial growth \(a(N) \sim N^g\), which means that imposing the quantum condition of \(\setZ_N\) symmetry reduces the number of possibilities drastically from the exponential growth (in \(N\)) for the number of generic \(N\)--covers.

\bigskip

Next we need to specify a quiver. This is done by  fixing one possible pants 
decomposition of the Riemann surface in the base; for a given \(\Sigma_{g,0}\), different pants decompositions give rise to S--dual phases that have to be studied separately. This choice distinguishes the electric and the magnetic lines.
The former are represented in $\Sigma_{g,0}$ by oriented closed \M2 lines along the cut, and the latter are the dual cycles in $\Sigma_{g,0}$.
Following~\cite{Amariti:2015dxa}, a given \(N\)--cover \(\Sigma^N_{g,0}\) fixes the allowed charges in the gauge theory because we only allow lines that are closed geodesics in the cover.
From a purely geometric perspective, an important difference is that the fundamental group is non-Abelian and the number of its generators grows with the order \(N\) of the cover. While in the torus case, we could readily identify \(\pi_1(T_{N;k,i})\) with the charge lattice, here we need to define a projection. We adopt the following strategy. Consider all the closed curves in the \(N\)--cover (distinguished by their holomogy class in \(\Sigma^N_{g,0}\)), project them on the base, identifying two curves that differ by an adjoint action \(\mathcal{C} \sim \Ad_x \mathcal{C} = x \mathcal{C} x^{-1}\). The electric and magnetic charges of the corresponding \ac{bps} operator are given by the projection of this curve on the cuts and their dual cycles on the base.

\begin{figure}
  \centering
  \begin{tikzpicture}
    \node at (0,0) {\includegraphics[width=5cm]{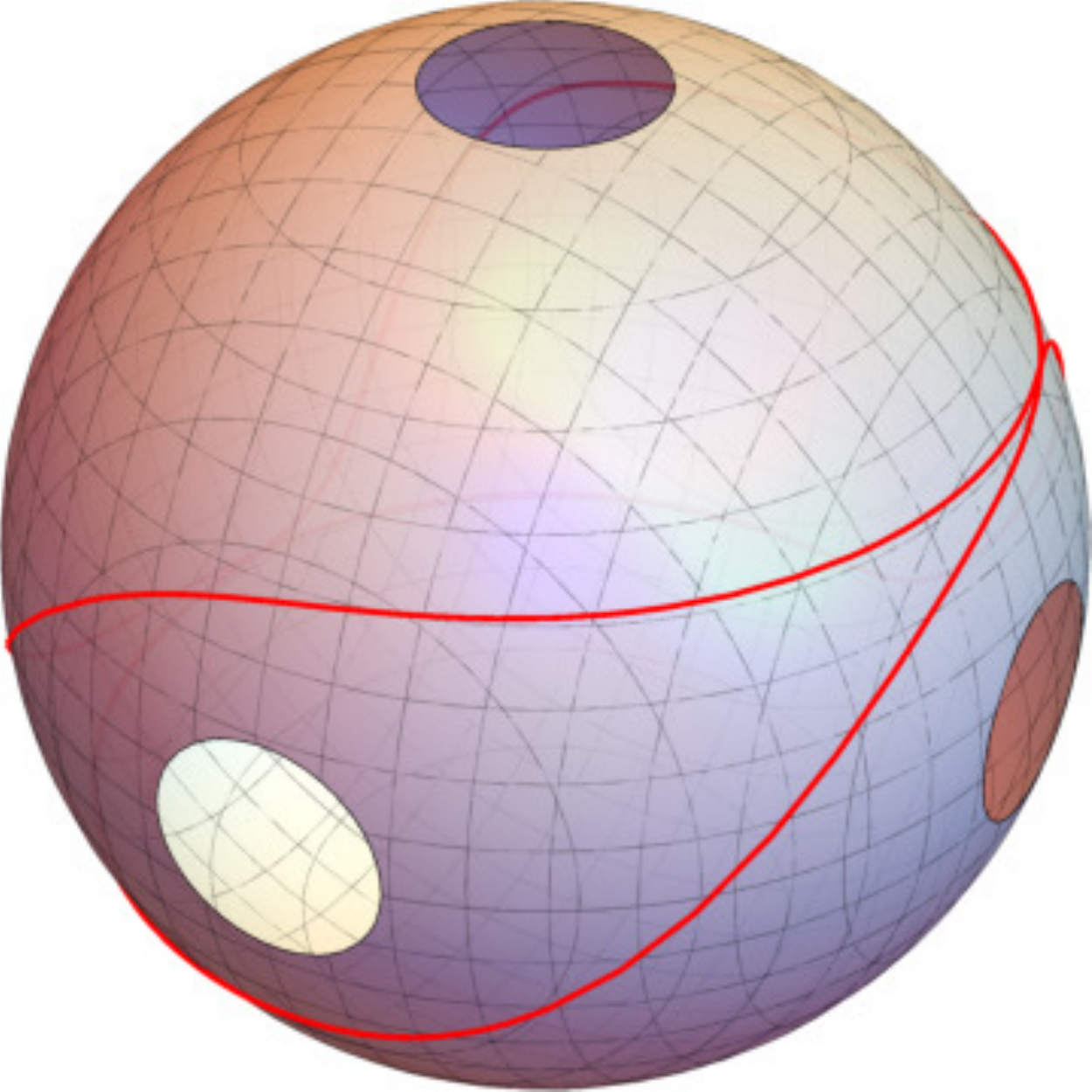}};
  \end{tikzpicture}
  \caption{A pair of pants. A generic line must satisfy
    \(n_1 [p_1] + n_2 [p_2] + n_3 [p_3] = 0 \), where \(p_i\) are the
    punctures.}
  \label{fig:pair-of-pants}
\end{figure}

This is however not enough. The quantum constraint discussed above plays a non-trivial role also for the electric lines. The allowed electric lines have to respect the quantum condition
on each pair of pants, \emph{i.e.} we keep only those electric lines that respect, on each pair
of pants, the $\mathbb{Z}_N$  symmetry. Geometrically speaking, a line operator for the theory  is only allowed if the corresponding curve is not broken by the pants decomposition, \emph{i.e.} if it satisfies the condition
\begin{equation}
  \label{eq:consistency-in-pants}
  n_1 [p_1] + n_2 [p_2] +n_3 [p_3]=0,
\end{equation}
where the \(p_i\) are the punctures of the surface \(\Sigma_{0,3}\) (see
Figure~\ref{fig:pair-of-pants}).
In the typical configuration, the numbers \(n_1\) and \(n_3\) are fixed by the topology of the cover to be integer multiples of fixed parameters \(k\) and \(k'\). In this case, using the fact that \([p_2] = -[p_1] - [p_3]\), the condition above implies that
\begin{equation}
  \begin{cases}
    n_2 = m_1 k & m_1 \in \setZ\\
    n_2 = m_3 k' & m_3 \in \setZ
  \end{cases}
\end{equation}
which means that \(n_2 \) must be an integer multiple of the least common multiple of \(k\) and \(k'\):
\begin{equation}
  n_2 = m_2 \lcm(k,k').
\end{equation}
Since the pants decomposition cuts through all the magnetic base cycles, the allowed magnetic lines are found by considering the whole surface without cuts, and imposing 
$\mathbb{Z}_N$ on the full quiver.

\bigskip

Repeating the construction for all the allowed \(N\)--covers of the surface \(\Sigma_{g,0}\) and projecting the lines on all the possible pants decompositions, we obtain the full classification of the class S theories with algebra \(A_{N-1}\) based on these surfaces.

\bigskip

We conclude this section by stressing an important result.
The mutual locality of the four-dimensional lines has not been imposed here, but it has been obtained as
a bonus of this construction.
This generalizes the result of~\cite{Amariti:2015dxa} for the case of $\mathcal{N}=4$ \ac{sym}.
Here, nevertheless,  there is a \emph{caveat} in this derivation: we have obtained the generalization of the \ac{dsz} quantization condition by taking into account the quantum constraints arising in six dimensions.
This constraint, as explained above, can be reformulated as the presence of a global $\mathbb{Z}_N$
symmetry in four dimensions~\cite{Gaiotto:2009we,Gadde:2011ik}.
Observe that by imposing this constraint, we exclude some of the multiple covers which are allowed 
by the geometry.
We can however work out the quantization condition also for these theories.
They can be interpreted as theories in which (some of) the $A_{N-1}$ are partially broken to 
subalgebras. We need to stress that also in those cases, a natural mutual locality condition 
is obtained for the line operators by the same construction discussed above.

\section{Example: genus 2}
\label{sec:genus2}

In this section, we provide a detailed analysis of the simplest class S theory, defined by an \M5--brane doubly
wrapped on a genus 2 compact surface.
This is the simplest example for a multiple cover of a genus $g>1$ Riemann surface, but it illustrates all the salient points of the procedure.

Consider two \M5--branes wrapping a genus two Riemann surface without punctures \(\Sigma_{2,0}\) (double torus). The fundamental group of the surface has four generators and one relation and admits the presentation
\begin{equation}
  \pi_1(\Sigma_{2,0}) = \braket{a_1, b_1, a_2, b_2}{\comm{a_1}{b_1} \comm{a_2}{b_2}=e },
\end{equation}
where
\begin{equation}
  \comm{a}{b} = a b a^{-1} b^{-1}.
\end{equation}

According to Eq.~(\ref{eq:newgenus}), a double cover of \(\Sigma_{2,0}\) is a Riemann surface of genus \(g' = N (g -1 ) + 1 = 2 (2 - 1) + 1 =3\) whose fundamental group \(\pi_1(\Sigma^2_{2,0})\) is a subgroup of \(\pi_1(\Sigma_{2,0})\). There are fifteen such subgroups, labeled by two ordered pairs of partitions of \(2\), \emph{i.e.} one partition per generator of the base \(\Sigma_{2,0}\). We collect them in Table~\ref{tab:fundamental-group-subgroups}, omitting the relation among the generators that is still the same as for the double torus in the base, namely
\begin{equation}
  \comm{a_1}{b_1} \comm{a_2}{b_2} = e.
\end{equation}

\begin{table}
  \centering
  \begin{tabular}{lcc}
    \toprule
    generators                                                                         & \((a_1, b_1)\)                       & \((a_2, b_2)\)                       \\ \midrule
    \(\set{a_1^2, b_1, \comm{a_2}{b_2}, a_1a_2, \Ad_{a_1} a_2^2, \Ad_{a_1}b_2} \)      & \((\ydiagram{2},\ydiagram{1,1})\)    & \((\ydiagram{2},\ydiagram{1,1})\)    \\[1em]
    \(\set{a_1, b_1^2, \comm{a_2}{b_2}, b_1a_2, \Ad_{b_1} a_2^2, \Ad_{b_1}b_2} \)      & \((\ydiagram{1,1},\ydiagram{2})\)    & \((\ydiagram{2},\ydiagram{1,1})\)    \\[1em]
    \(\set{a_1^2, a_1b_1, \comm{a_2}{b_2}, a_1a_2, \Ad_{a_1} a_2^2, \Ad_{a_1}b_2} \)   & \((\ydiagram{2},\ydiagram{2})\)      & \((\ydiagram{2},\ydiagram{1,1})\)    \\[1em]
    \(\set{a_1^2, b_1, \comm{a_2}{b_2}, a_1b_2, \Ad_{a_1} b_2^2, \Ad_{a_1}a_2} \)      & \((\ydiagram{2},\ydiagram{1,1})\)    & \((\ydiagram{1,1},\ydiagram{2})\)    \\[1em]
    \(\set{a_1, b_1^2, \comm{a_2}{b_2}, b_1b_2, \Ad_{b_1} b_2^2, \Ad_{b_1}a_2} \)      & \((\ydiagram{1,1},\ydiagram{2})\)    & \((\ydiagram{1,1},\ydiagram{2})\)    \\[1em]
    \(\set{a_1^2, a_1b_1, \comm{a_2}{b_2}, a_1b_2, \Ad_{a_1}b_2^2, \Ad_{a_1}a_2} \)    & \((\ydiagram{2},\ydiagram{2})\)      & \((\ydiagram{1,1},\ydiagram{2})\)    \\[1em]
    \(\set{a_1^2, b_1, \comm{a_2}{b_2}, a_1a_2, \Ad_{a_1} a_2^2, \Ad_{a_1}a_2b_2} \)   & \((\ydiagram{2},\ydiagram{1,1})\)    & \((\ydiagram{2},\ydiagram{2})\)      \\[1em]
    \(\set{a_1, b_1^2, \comm{a_2}{b_2}, b_1a_2, \Ad_{b_1} a_2^2, \Ad_{b_1}a_2b_2} \)   & \((\ydiagram{1,1},\ydiagram{2})\)    & \((\ydiagram{2},\ydiagram{2})\)      \\[1em]
    \(\set{a_1^2, a_1b_1, \comm{a_2}{b_2}, a_1a_2, \Ad_{a_1}a_2^2, \Ad_{a_1}a_2b_2} \) & \((\ydiagram{2},\ydiagram{2})\)      & \((\ydiagram{2},\ydiagram{2})\)      \\[1em]
    \(\set{a_1^2, b_1, a_2, b_2, \Ad_{a_1}a_2, \Ad_{a_1} b_2}\)                        & \((\ydiagram{2}, \ydiagram{1,1})\)   & \((\ydiagram{1,1}, \ydiagram{1,1})\) \\[1em]
    \(\set{a_1, b_1^2, a_2, b_2, \Ad_{b_1}a_2, \Ad_{b_1} b_2}\)                        & \((\ydiagram{1,1}, \ydiagram{2})\)   & \((\ydiagram{1,1}, \ydiagram{1,1})\) \\[1em]
    \(\set{a_2^2, b_2, a_1, b_1, \Ad_{a_2}a_1, \Ad_{a_2} b_1}\)                        & \((\ydiagram{1,1}, \ydiagram{1,1})\) & \((\ydiagram{2}, \ydiagram{1,1})\)   \\[1em]
    \(\set{a_2, b_2^2, a_1, b_1, \Ad_{b_2}a_1, \Ad_{b_2} b_1}\)                        & \((\ydiagram{1,1}, \ydiagram{1,1})\) & \((\ydiagram{1,1}, \ydiagram{2})\)   \\[1em]
    \(\set{a_1^2, a_1 b_1, a_2, b_2, \Ad_{a_1}a_2, \Ad_{a_1} b_2}\)                    & \((\ydiagram{2}, \ydiagram{2})\)     & \((\ydiagram{1,1}, \ydiagram{1,1})\) \\[1em]
    \(\set{a_2^2, a_2 b_2, a_1, b_1, \Ad_{a_2}a_1, \Ad_{a_2} b_1}\)                    & \((\ydiagram{1,1}, \ydiagram{1,1})\) & \((\ydiagram{2}, \ydiagram{2})\)     \\
    \bottomrule
  \end{tabular}
  \caption{Generators of the subgroups of the fundamental group of the double torus. Each subgroup is labeled by two ordered pairs of partitions of \(2\), here represented by Young diagrams.}
  \label{tab:fundamental-group-subgroups}
\end{table}

\begin{figure}
  \centering
  \includegraphics[width=8cm]{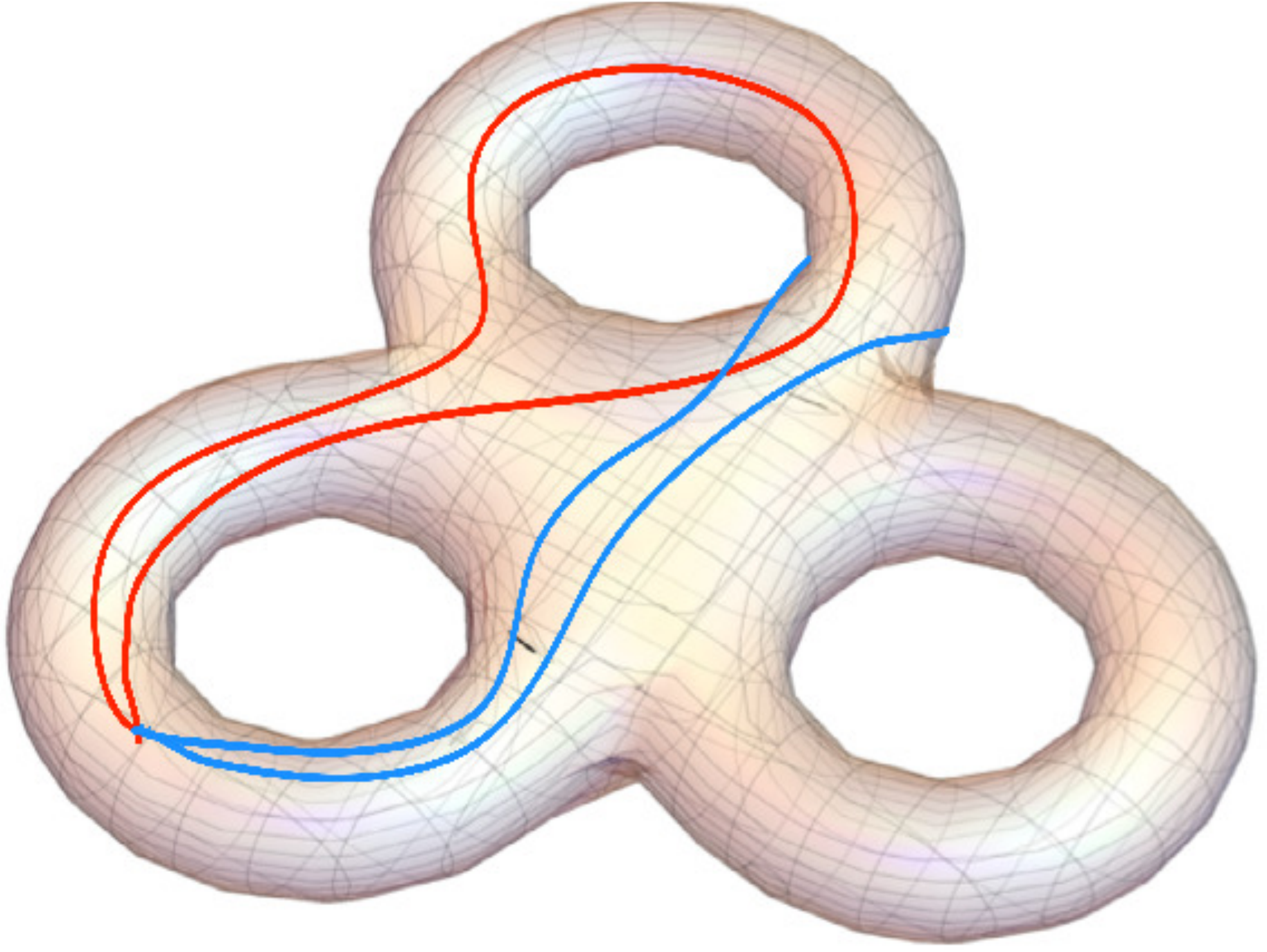}
  \caption{The double cover \(\Sigma^2_{2,0}\) with fundamental group generated by \(\{ a_1, b_1^2, a_2, b_2, \Ad_{b_1}a_2, \Ad_{b_1} b_2 \}\). The cover is associated to the partitions \(\set{(\ydiagram{2},\ydiagram{1,1}),(\ydiagram{1,1},\ydiagram{1,1})}\). The pair \((\ydiagram{1,1},\ydiagram{1,1})\) indicates that the punctured torus in the base with generators \((a_2,b_2)\) is lifted to a disjoint cover. The cycles \(\Ad_{b_1}a_2, \Ad_{b_1} b_2\) are drawn respectively in blue and red.}
  \label{fig:non-acceptable-cover}
\end{figure}

As mentioned before, we need to distinguish two classes. In the first class (the first nine covers in Table~\ref{tab:fundamental-group-subgroups}), at least one of the partitions in each pair is of the type \(\ydiagram{2}\) (transposition of two elements); in the other (the last six covers in Table~\ref{tab:fundamental-group-subgroups}), both the partitions associated to one of the pairs \((a_i, b_i)\) are of the type \(\ydiagram{1,1}\) (the identity permutation).
Covers of the second class are excluded as they do not respect the quantum condition. In the pants decomposition of \(\Sigma_{2,0}\) into two one-punctured tori, the pair of partitions \((\ydiagram{1,1}, \ydiagram{1,1})\) means that in the corresponding double cover, there is a non-maximal puncture (see Figure~\ref{fig:double-covers-double-torus}). In Figure~\ref{fig:non-acceptable-cover}, we show an explicit example where one of the two tori with one puncture is lifted to a disconnected double cover, indicating a non-maximal puncture.
This leaves us with nine allowed double covers. Each of these corresponds to a class S quiver with algebra \(A_1 \oplus A_1 \oplus A_1\). We can read off the global group structure from the fundamental group of \(\Sigma^2_{2,0}\). Take for example the cover with \(\pi_1\) given by
\begin{equation}
  \pi_1 (\Sigma^2_{2,0}) = \braket{a_1^2, b_1, \comm{a_2}{b_2}, a_1 a_2, \Ad_{a_1} a_2^2, \Ad_{a_1}b_2}{\comm{a_1}{b_1} \comm{a_2}{b_2}=e}.
\end{equation}
The generators have been chosen such that the symplectic form is given by three copies of \(\begin{psmallmatrix} 0 & 1 \\ -1 & 0 \end{psmallmatrix}\). This means that two closed lines \(\mathcal{C}\) and \(\mathcal{C}'\) on the double cover with homologies
\begin{equation}
  \begin{aligned}
    [\mathcal{C}] &= p_1 [a_1^2] + q_1 [b_1] + p_2 [\comm{a_2}{b_2}] + q_2 [a_1 a_2] + p_3 [\Ad_{a_1} a_2^2] + q_3 [ \Ad_{a_1}b_2], \\
    [\mathcal{C}'] &= p_1' [a_1^2] + q_1' [b_1] + p_2' [\comm{a_2}{b_2}] + q_2' [a_1 a_2] + p_3' [\Ad_{a_1} a_2^2] + q_3' [ \Ad_{a_1}b_2],
  \end{aligned}
\end{equation}
will intersect
\begin{equation}
  \label{eq:intersection-in-the-cover}
  \braket{\mathcal{C}}{\mathcal{C}'} = p_1 q_1' - p_1' q_1 + p_2 q_2' - p_2' q_2  + p_3 q_3' - p_3' q_3  
  \in \mathbb{Z}
\end{equation}
times.

\bigskip

So far, we have used only topological data (the homology in the double cover). We need to express it in terms of the allowed charges in the quiver gauge theory. As in~\cite{Amariti:2015dxa}, we can probe the \M5--brane geometry with \M2--branes extended in the directions \(x^0, x^4\) and wrapping a \emph{closed finite length geodesic curve on the double cover}. If we reduce to \tIIB, such a brane turns into a \D1--\F1 bound state which corresponds to a \ac{bps} state in the gauge theory. 

\begin{figure}
  \centering
  \begin{tikzpicture}
    \node at (-4,1) {\includegraphics[width=5cm]{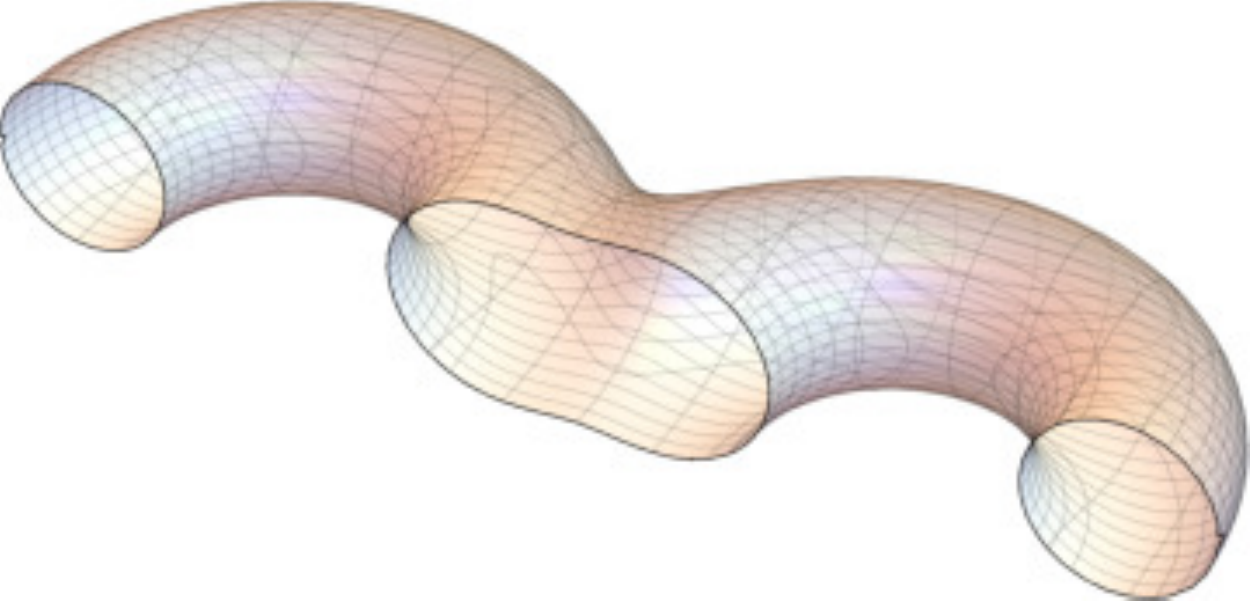}};
    \node at (-5,-1) {\includegraphics[width=5cm]{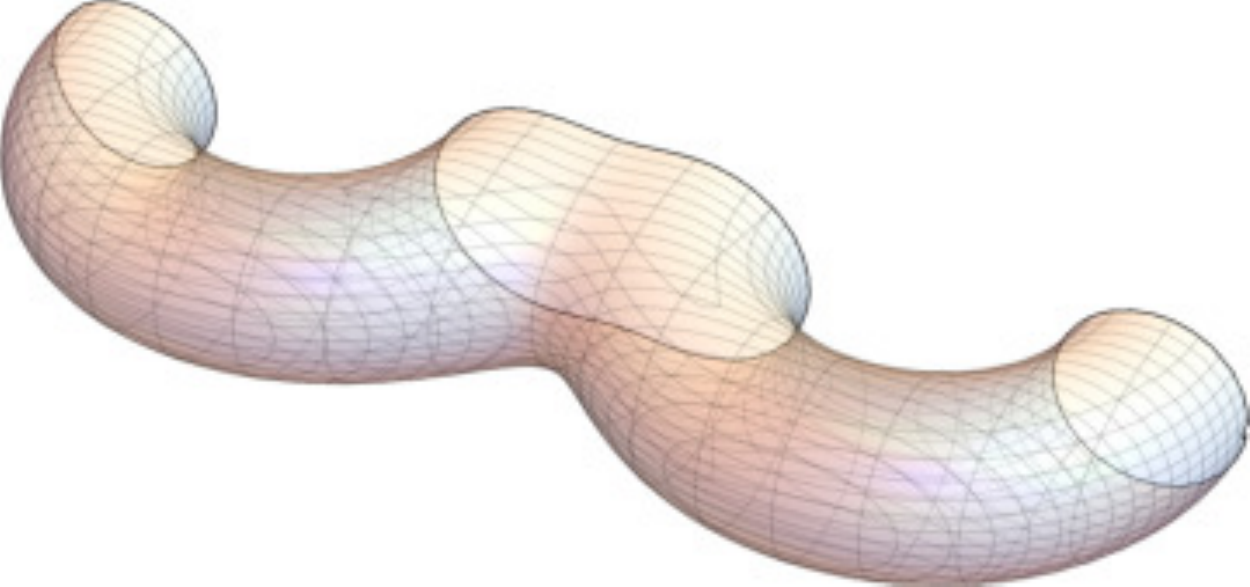}};
    \node at (-4,-3){(a)};
    \node at (1,0) {\includegraphics[width=3cm]{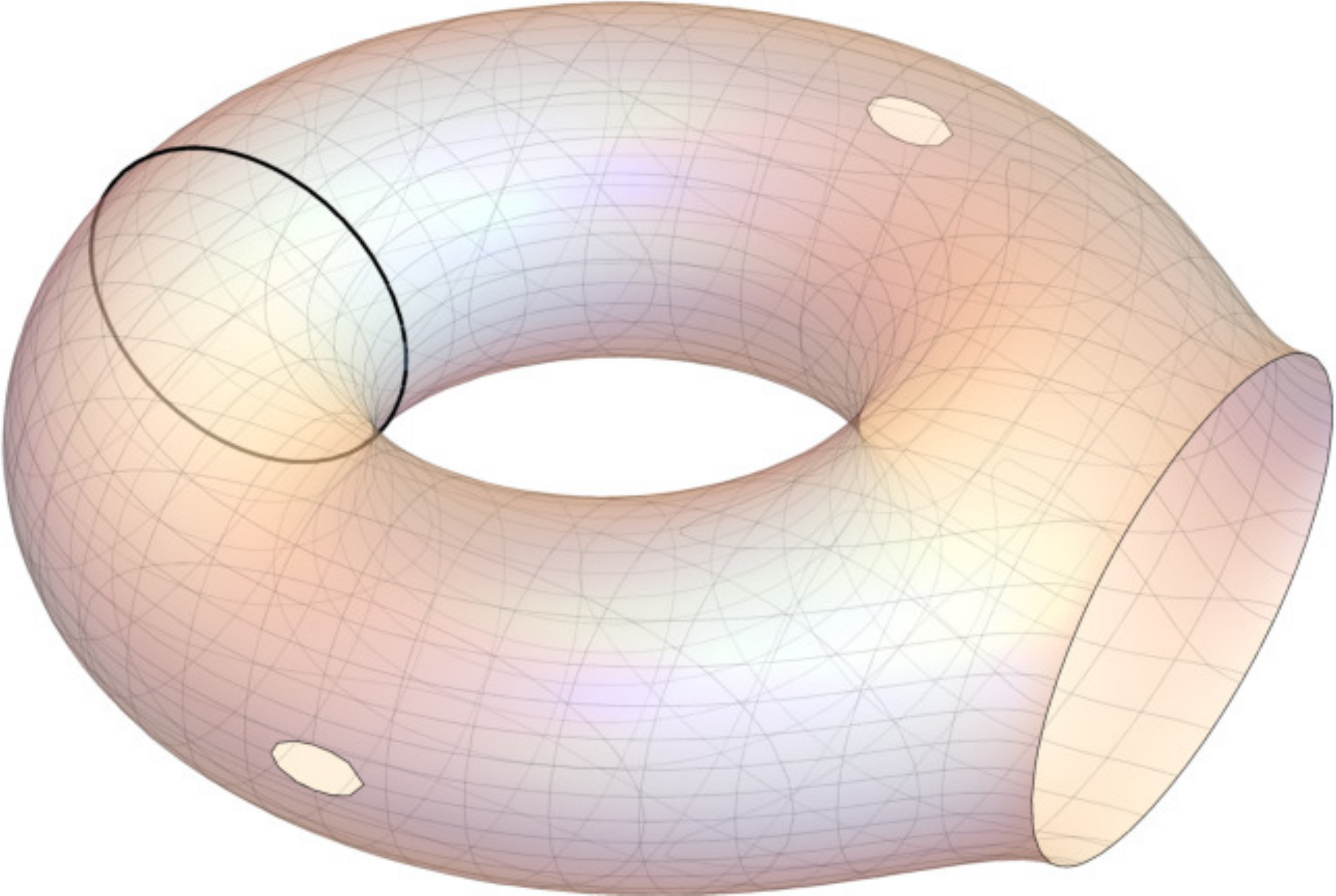}};
    \node at (4,-1) {\includegraphics[width=3cm]{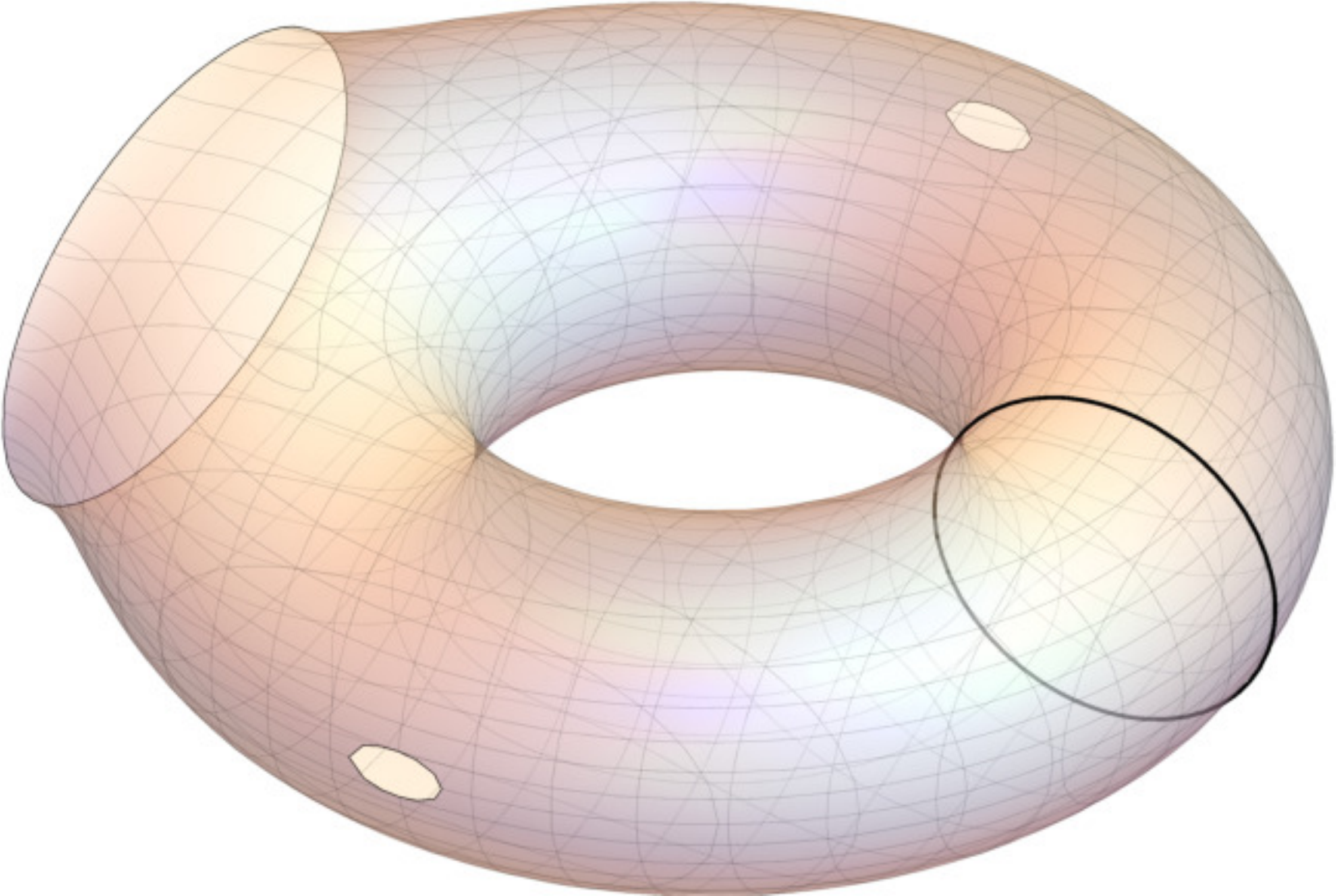}};
    \node at (2,-3){(b)};
  \end{tikzpicture}
  \caption{Pant decomposition of the double torus in the base along the cycles \(a_1\), \(a_1 a_2\) and \(a_2\) (a) and the cycles \(a_1\), \(\comm{a_1}{b_1}\), \(a_2\) (b)}
  \label{fig:double-torus-pant-decomposition-1}
\end{figure}

For a given double cover \(\Sigma^2_{2,0}\) and \M2--brane wrapping a fixed curve \(\mathcal{C}\), there exist multiple possible S--dual interpretations that correspond to pants decompositions of the Riemann surface \(\Sigma_{2,0}\) in the base. Let us consider one such decomposition, in which we cut the double torus along the cycles \(a_1\), \(a_1 a_2\) and \(a_2\) as in Figure~\ref{fig:double-torus-pant-decomposition-1}(a).
Following~\cite{Gaiotto:2009we}, each cut identifies a gauge group. A line in the same homology class as a given cut is a Wilson line for the corresponding group and its homology is the corresponding electric charge. While this is enough in the case of the torus \(\Sigma_{1,0}\) which has an Abelian fundamental group, higher-genus Riemann surfaces have lines with trivial homology that are nevertheless non-contractible, such as the curve \(\comm{a_1}{b_1}\). In this case, we adopt the following strategy. Curves are distinguished by their homology class on the multiple cover. We project the curve \(\mathcal{C}\) living on the double cover \(\Sigma^2_{2,0}\) on the base, identifying curves that differ by an adjoint action \(\mathcal{C} \sim \Ad_x \mathcal{C} \), and then we project the curve on the cuts to read off the corresponding electric charges. The same procedure is repeated to obtain the magnetic charges, which are defined with respect to the dual cycles, which in this choice of pants are \(\set{b_1, \comm{a_1}{b_1}, b_2}\). 
Concretely, a line \(\mathcal{C}\) with homology \( [\mathcal{C}] = p_1 [a_1^2] + q_1 [b_1] + p_2 [a_1 a_2] + q_2 [\comm{a_2}{b_2}] + p_3 [\Ad_{a_1} a_2^2] + q_3 [ \Ad_{a_1}b_2]\) corresponds to a \ac{bps} state with charges \((2p_1, q_1)\), \((p_2, q_2)\), \((2p_3, q_3)\).
Not all such states are compatible with the gauge theory requirement that the global symmetry in each pair of pants is \(\setZ_2\) as opposed to \(\setZ_2^3\), and we need to impose the consistency condition in Eq.~\eqref{eq:consistency-in-pants}. Take for example the upper pair of pants in Figure~\ref{fig:double-torus-pant-decomposition-1}(a). The three punctures are \(p_1 = a_1\), \(p_2 = (a_1 a_2)^{-1}\) and \(p_3 = a_2\), hence we have the condition
\begin{equation}
  n_1 [a_1] + n_2 [(a_1 a_2)^{-1}] + n_3 [a_2] = 0,
\end{equation}
which implies \(n_1 = n_2 = n_3\).  From the geometry of the double cover, we know that \(n_1 = 2 p_1\) and \(n_3 = 2 p_3\), which means that the only acceptable line operators are those for which \(p_2 \) is also even. Rewriting \(p_1 = e_1\), \(p_2 = 2 e_2\), \(p_3 = e_3\) and \(q_i = m_i\), we can rewrite the intersection number in Eq.~\eqref{eq:intersection-in-the-cover} in terms of charges and find the expected \ac{dsz} quantization condition for \(A_1 \oplus A_1 \oplus A_1\):
\begin{equation}
  \braket{\mathcal{C}}{\mathcal{C}'} = 2 (e_1 m_1' - e_1' m_1 ) + 2 (e_2 m_2' - e_2' m_2 ) + 2 (e_3 m_3' - e_3' m_3 ) \in \setZ .
\end{equation}
Note that the pants decomposition only preserves information about the electric groups since it breaks all 't~Hooft cycles. Consistency conditions for the magnetic lines can only be imposed when considering the full Riemann surface.

\bigskip

We can now rewrite all the information obtained from the double cover in terms of the allowed charge lattices for the three groups in the quiver gauge theory corresponding to the pants decomposition in Figure~\ref{fig:double-torus-pant-decomposition-1}(a). In the notation of~\cite{Amariti:2015dxa}, the three lattices associated to the cuts \(a_1\), \((a_1 a_2)^{-1}\) and \(a_2\) are, respectively,
\begin{align}
  \Gamma_{2;2,0} &= \begin{pmatrix}
    2 & 0 \\ 0 & 1
  \end{pmatrix}, & 
                  \Gamma_{2;2,0} &= \begin{pmatrix}
                    2 & 0 \\ 0 & 1
                  \end{pmatrix}, & 
                                  \Gamma_{2;2,0} &= \begin{pmatrix}
                                    2 & 0 \\ 0 & 1
                                  \end{pmatrix} ,
\end{align}
so that the gauge groups are \(SO(3)_+ \times SO(3)_+ \times SO(3)_+\).

The same construction can be repeated for the same double cover, but now choosing the alternative pants decomposition in Figure~\ref{fig:double-torus-pant-decomposition-1}(b). In this case, one finds that the three lattices associated to the cuts \(a_1\), \(\comm{a_1}{b_1}\) and \(a_2\) are, respectively,
\begin{align}
  \Gamma_{2;2,0} &= \begin{pmatrix}
    2 & 0 \\ 0 & 1
  \end{pmatrix}, & 
                  \Gamma_{2;1,0} &= \begin{pmatrix}
                    1 & 0 \\ 0 & 2
                  \end{pmatrix}, & 
                                  \Gamma_{2;2,0} &= \begin{pmatrix}
                                    2 & 0 \\ 0 & 1
                                  \end{pmatrix} ,
\end{align}
so that the gauge groups are now \(SO(3)_+ \times SU(2) \times SO(3)_+\). The two quivers associated to the same double cover \(\Sigma^2_{2,0}\) are found to be S--dual to each other.

\bigskip

The construction can be repeated for all the nine allowed double covers. One can verify that two of the three lattices are fixed by the choice of the cover and are labelled by pairs of permutations as follows:
\begin{equation}
  \begin{aligned}
    (\ydiagram{2},\ydiagram{1,1}) &\mapsto   \Gamma_{2;2,0} = \begin{pmatrix}  2 & 0 \\ 0 & 1 \end{pmatrix}, &
    (\ydiagram{1,1},\ydiagram{2}) &\mapsto   \Gamma_{2;1,0} = \begin{pmatrix}  1 & 0 \\ 0 & 2 \end{pmatrix}, \\
    (\ydiagram{2},\ydiagram{2}) &\mapsto   \Gamma_{2;2,1} = \begin{pmatrix}  2 & 0 \\ 1 & 1 \end{pmatrix} .
  \end{aligned}
\end{equation}
The third lattice depends on the pants decomposition. It is always \(\Gamma_{2;1,0}\) for the choice in Figure~\ref{fig:double-torus-pant-decomposition-1}(b) and it is \(\Gamma_{2;2,0}\) for the choice in Figure~\ref{fig:double-torus-pant-decomposition-1}(a) for all the double covers but \(\set{(\ydiagram{1,1},\ydiagram{2}), (\ydiagram{1,1},\ydiagram{2}) }\), where we find that the lattice associated to the cut \((a_1 a_2)^{-1}\) is \(\Gamma_{2;1,0}\).

This concludes the classification of all the \(T^2\) theories that can be obtained starting from a double torus \(\Sigma_{2,0}\).

\bigskip
It is interesting to see what happens when the same procedure is
applied to one of the excluded covers. Take for example the one in
Figure~\ref{fig:non-acceptable-cover}, with generators \(\set{a_1, b_1^2, a_2, b_2, \Ad_{b_1}a_2, \Ad_{b_1} b_2}\). If we pick one of the pants decompositions in Figure~\ref{fig:double-torus-pant-decomposition-1}, the projection of the closed lines in the cover on the base gives two lattices. In the first, all electric charges and even-valued magnetic charges are allowed, while in the second, all magnetic and electric charges are possible. In other words, the intersection number of two closed lines in the cover is projected to the following condition on the \ac{bps} charges:
\begin{equation}
  \braket{\mathcal{C}}{\mathcal{C}'} =  2 \pqty{e_1 m_1' - e_1' m_1} + \pqty{e_2 m_2' - e_2' m_2} \in \setZ ,
\end{equation}
which is consistent with a \ac{dsz} quantization for the algebra corresponding to \(SU(2) \times U(1)\).
This discussion may have some relevance for the generalization of our result in presence of (non-maximal) punctures.
We will return to this issue in the conclusions.

\section{Generalizations}
\label{sec:generalizations}

In the previous example, we have described in detail the construction for \(g = 2\) and \(N = 2\). Here we discuss the new features that arise in the general case.

Let us start with \(g = 2\), \(N > 2\). According to the Riemann--Hurwitz theorem, the genus of the cover grows linearly with the order as \(g' = N + 1\). This means that each time we increase the order of the cover, there are two more generators in the fundamental group of \(\Sigma^N_{g,0}\) that have to be projected on the base. It turns out that these new generators do not change the description on the base. In the case of the allowed covers, all the new generators can be chosen in the same conjugacy class as \(\comm{c}{d}\) and \(ac\). Since we identify two curves that differ only by an adjoint action on the base, they do not add any new information but contribute only to the charges of the group identified by the middle cut. After the identification, the projection works like in the double cover and the
generalization is straightforward. For illustration, the fundamental group \(\pi_1(\Sigma^3_{2,0})\) of an allowed 
triple cover of the double torus is generated by
\begin{equation}
  \set{a^3, b, \comm{c}{d}, a c, \Ad_{ac^{-1}} c^3, \Ad_{ac^{-1}} d, \Ad_{ac^{-1}} \comm{c}{d}, \Ad_{a} ac } .
\end{equation}
In the pants decomposition of
Figure~\ref{fig:double-torus-pant-decomposition-1}(a), this cover corresponds to three copies of the lattice \(\Gamma_{3;3,0}\), \emph{i.e.} to the quiver \(((SU(3)/\setZ_3)_0)^3\).

\bigskip

The second generalization that we need to address is $g>2$.
In this case, there are $g$  fundamental cycles generating $\Sigma_{g,0}$.
There are also $2(g-1)$ pairs of pants and  $3(g-1)$ gauge groups, corresponding to the number of gluings.
One can see that \(g -2 \) of these cuts are topologically equivalent, which leaves us with \(3(g-1) - (g -2) = 2 g - 1\) independent charge lattices.
By the argument above, we just need to look at \(N = 2\), since covers
of higher order will just result in equivalent data. The genus of
\(\Sigma^2_{g,0}\) is \(2 ( g - 1) + 1 = 2g -1\). The fundamental
group has thus precisely the right number of generators necessary to
project the homologies of the lines in the multiple cover on the
charge lattice in the quiver gauge theory. 
At this point, we need to impose the consistency condition on each pair of pants, as we have done 
in the case of the double cover.
After this quantum condition is imposed, the
$\mathbb{Z}_g^2$ lattice is obtained.

In summary, the charge lattice of a class S theory obtained by the partially twisted compactification 
of the $A_{N-1}$ $\mathcal{N}=(2,0)$ theory on $\Sigma_{g,0}$ can be worked out as follows. 
First, consider all the possible $N$--covers $\{$ \(\Sigma_{g,0}^N\) $\}$  by 
listing the possible realizations of each
$\pi_1( \Sigma_{g,0}^N ) $ in terms of the fundamental cycles generating $\Sigma_{g,0}$.
Then use the quantum condition~\cite{Gaiotto:2009we,Gadde:2011ik} to exclude the covers
having one pair of dual cycles associated to the identity 
permutation.
At this point, we are left with 
a lattice of dimension $\mathbb{Z}_{2g-1}^2$, as some of the new cycles in the cover are identified with some of the fundamental cycles in $\Sigma_{g,0}$.
Let us conclude with a general comment: the maximality of the charge lattice is automatic here.
The reason is essentially the same as in  \cite{Amariti:2015dxa}:
a theory containing \ac{WH} lines in the adjoint  is inconsistent because it
correspond to an $N^2$--cover of $\Sigma_{g,0}$.

\section{Conclusions and further directions}
\label{sec:conclusions}

In this paper, we have shown how to specify the gauge group 
of class S theories on compact Riemann surfaces via an M--theory construction.
This was done by 
giving a prescription for the derivation of  the charge lattice of the \ac{WH} line operators.
We have shown that the lattices can be extracted in M--theory by probing the
\M5--branes wrapping the compact Riemann surface with \M2--lines.
As in the case of $\mathcal{N}=4$ \ac{sym}, the information of the lattices is encoded in the fundamental group of another Riemann 
surface obtained from the multiple wrapping, namely the $N$--cover of the Riemann surface.
We found that in the case of class S theories, differently from $\mathcal{N}=4$ \ac{sym}, the quantum properties of the six-dimensional 
theory play a role.\footnote{A similar discussion has appeared 
in~\cite{Tachikawa:2013hya,Xie:2013vfa}.}
We need to impose a quantum constraint, which has two effects: first, it selects only some of the 
$N$--covers as corresponding to acceptable quantum theories. This is due to the global $\mathbb{Z}_{N}$
symmetry left in four dimensions. 
This projection turns the exponential growth of the number of allowed covers
into a polynomial one.
Second, it imposes constraints on the allowed set of charges.
The latter follows from considering the pants decomposition of the theories
in terms of $T_N$ blocks and studying the role of the quantum constraint when gluing 
the blocks back together.
We find that by studying the projection of the homologies of the \M2--lines from the $N$--cover geometry to the base Riemann surface and by imposing the quantum constraints properly, we can obtain the charge lattice of the class S theories and
consequently the gauge group.

Our analysis has a counterpart in the discussion of~\cite{Tachikawa:2013hya}.
The main observation behind~\cite{Tachikawa:2013hya} is that the six-dimensional $\mathcal{N}=(2,0)$ theories
with non-simple gauge algebra are  non-conventional quantum field theories.
 They are usually referred to as relative field theories, specified by their gauge algebra rather than their gauge group~\cite{Freed:2012bs}.
Additional topological data is required for describing these theories on curved manifolds.
In other words, such a six-dimensional theory does not have a partition function but a partition vector~\cite{Witten:2009at}. 
By considering the six-dimensional theory on  a curved manifold (here a compact Riemann surface), one specifies a direction in the partition vector leading to the partition function of the four-dimensional theory.
With this choice, there is a $\mathbb{Z}_N$ leftover central symmetry,
originating the quantum constraint discussed in this paper. 
In the four-dimensional language, this choice of a partition function represents the additional data necessary to
specify the gauge group.
The origin of the lattices has also been studied in F--theory in~\cite{DelZotto:2015isa}.
In this case, there are tensionless strings associated to 
\D3--branes wrapping spheres in the base of the elliptically fibered CY$_3$.
In F--theory, the center of the gauge group is associated to the so called \emph{defect
group}, representing the mismatch 
between the charge lattice of those strings and the dual lattice.
The projection of the partition vector on the partition function in this case is done by specifying 
the choice of the background fluxes on the compact manifold.
It would be interesting to connect this discussion with our results.

The origin of the lattices can be also understood as the decoupling of
the $U(1)$ from the original $U(N)$ gauge symmetry of the 
theory living on the stack of $N$ \M5~branes.
A similar discussion  can be found in \cite{Moore:2014gua} 
for $\mathcal{N}=4$ \ac{sym} in a type \tIIB setup.
Here, for class S theories, we have studied the problem in terms of 
the topology of the multiple cover, and we have reformulated
the decoupling of the $U(1)$ in terms of the 
leftover $\mathbb{Z}_N$ symmetry. 
The two approaches lead to the same conclusions
\footnote{We thank Ofer Aharony for having drawn our attention to this relation.}.

A first natural generalization of our treatment is the case of Riemann surfaces with punctures. The construction should be simple, as these
punctures introduce only flavor symmetries and do not  heavily modify the 
structure of the gauge theories.
We do not need to consider maximal punctures.
In this case, the structure of the quantum constraints may differ from the
discussion above, which may lead to a generalization of the
quantum condition to impose on the $N$--covers obtained from the fundamental group. 

This observation is directly connected to the multiple covers that
do not respect the quantum condition. As mentioned above, the excluded covers are related to gauge theories where the gauge algebra is broken down to a subalgebra
of $A_{N-1}$ for the $N$--cover.
When gluing non-maximal punctures, this is related to a higgsing 
of the gauge group, and it is plausible that also those theories have to 
be considered. 
In this case, as discussed in Section~\ref{sec:genus2}, the expected mutual locality condition
follows from the projection of the homologies of the \M2 lines on the charge lattice.
This means that if the covers corresponding to such a higgsing are allowed in cases with non-maximal
punctures, then the mutual locality condition on the lines is automatically imposed.
The analysis of those cases requires further analysis.\footnote{We are grateful to Takuya Okuda for discussions on this point.}

One can also consider $\mathcal{N}=1$ descriptions of \M5--branes on
Riemann surfaces (see for example
\cite{Benini:2009mz,Bah:2011je}).
In this case, there are additional deformations
leading to a different gluing with $\mathcal{N}=1$.
Again, this difference may modify the quantum constraints, and understanding the global properties of those theories via our analysis is an
interesting problem.

A last question that naturally arises regards the possibility of having S--duality orbits. 
In $\mathcal{N}=4$ \ac{sym}, these orbits arise whenever $N$ is not square-free.
In our case, there is another ingredient in the S--duality group. This follows from 
having different S--dual quivers associated to different pants decompositions. 
This corresponds to the mapping class group of the Riemann surface. It would be interesting to study
the S--duality orbits in this case.

 \section*{Acknowledgments}
 The authors would like to thank Luis Alvarez--Gaumé, Claudius Klare, Takuya Okuda, Hugo Parlier, Alberto Zaffaroni, and Michele Del Zotto for enlightening discussions and comments, and Ofer Aharony for helpful comments on the manuscript. D.O. and S.R. would furthermore like to thank the \textsc{cern} Theory Department and the Department of Physics of Kyoto University for hospitality.

\appendix

\section{Mathematical appendix}
\label{sec:details}

In this appendix, we discuss the mathematical aspects of multiple covers of a Riemann surface
$\Sigma_{g,0}$.
We refer the reader to \cite{opac-b1122188} for review.
We start by introducing the notion of the fundamental group which encodes the closed curves on 
$\Sigma_{g,0}$. Then we introduce the notion of the multiple cover and 
associate it to the symmetric group of permutations $S_n$.
We also present some instructive examples, starting with 
the case of the free Abelian group, associated to the circle 
 $S^1$. Then we review the case of the torus, to fix the notation used extensively in the main body of the paper.

\subsection{The fundamental group}
\label{sec:fundamentalGroup}
Consider an orientable surface $\Sigma_{g,0}$.
 Two closed curves are 
 \emph{homotopic} if they can be continuously deformed into each other.
The \emph{fundamental group} is the set of homotopy classes of curves. The
product is defined as the composition of curves.

\begin{figure}
  \centering
  \begin{tikzpicture}
    \node at (-2,0) {\includegraphics[width=6cm]{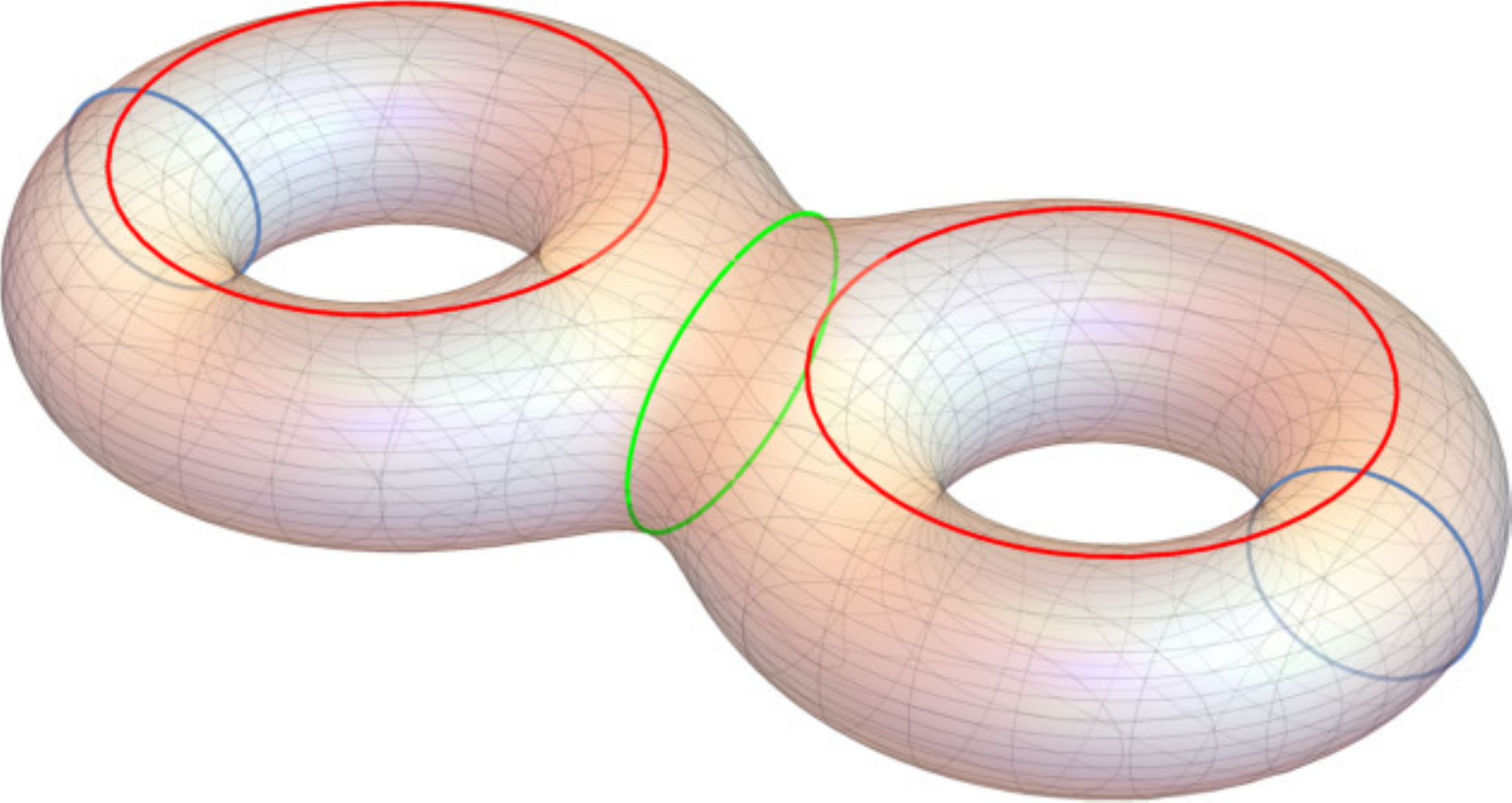}};

    \begin{scope}[shift={(5,0)},thick]
      \foreach \a/\b in {45/0, 90/135, 225/180, 270/315} {
        \draw[color=blue,-latex] (\a:2cm) -- (\b:2cm); }
      \foreach \a/\b in {90/45, 135/180, 270/225, 315/360} {
        \draw[color=red,-latex] (\a:2cm) -- (\b:2cm); }
    
      \draw[color=green,-latex] (270:2cm) -- (90:2cm);
    
      \node at (22.5:2.2cm) {\(a_1^{-1}\)}; 
      \node at (67.5:2.2cm) {\(b_1^{-1}\)}; 
      \node at (112.5:2.2cm) {\(a_2\)}; 
      \node at (157:2.2cm) {\(b_2\)}; 
      \node at (202.5:2.2cm) {\(a_2^{-1}\)};
      \node at (247.5:2.2cm) {\(b_2^{-1}\)}; 
      \node at (292.5:2.2cm) {\(a_1\)}; 
      \node at (337:2.2cm) {\(b_1\)};
      \node at (.8cm,0) {\(\comm{a_1}{b_1}\)};
    \end{scope}
  \end{tikzpicture}
  \caption{Double torus}
  \label{fig:double-torus}
\end{figure}
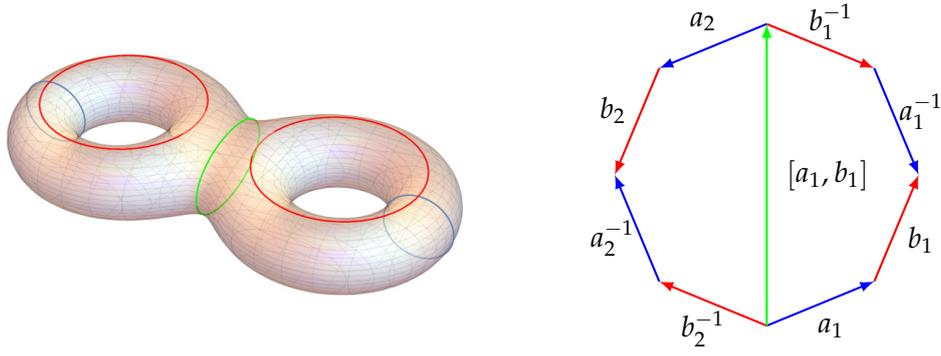

We can 
consider \emph{e.g.} the case of the compact genus one surface $\Sigma_{1,0}$, corresponding to the torus $T^2$.
We refer to this surface with both notations, hoping that it does not cause confusion.
This surface is generated by two cycles, $a$ and $b$, corresponding to 
two freely homotopic closed curves. The fundamental group is given by
\begin{equation}
  \pi_1(T^2) = \braket{a, b}{a b a^{-1} b^{-1} \equiv \comm{a}{b} = e} = \setZ \times \setZ,
\end{equation}
where $e$ is the \emph{identity}, \emph{i.e.} a cycle contractible to a point in the geometry.
In this case, the fundamental group of the torus $\pi_1(T^2)$ is Abelian and isomorphic to $\mathbb{Z} \times \mathbb{Z}$. 
Note that the torus can be represented as a square with edges $a$, $b$, $a^{-1}$ and $b^{-1}$.
In general, a compact genus-$g$ surface can be represented
by a $4g$-gon in which the edges corresponding to the cycles $a_i,\, a_i^{-1}$ and $b_i,\, b_i^{-1}$ are identified. The fundamental group is then defined as the group of \(2g \) generators and one relation
\begin{equation}
  \label{fundg}
  \pi_1(\Sigma_{g,0}) = \braket{a_j, b_j}{\prod_{j=1}^{g} \comm{a_j}{b_j} = e },
\end{equation}
where $a_i$ and $b_i$ are freely homotopic closed curves (see Figure~\ref{fig:double-torus} for \(\Sigma_{2,0}\)).

If the surface is orientable but non-compact (\emph{i.e.} it contains $n$ punctures 
$p_1,\dots,p_n$), 
the fundamental group in Eq.~(\ref{fundg}) is modified to
\begin{equation}
  \label{fundgpunct}
  \pi_1(\Sigma_{g,n}) = \braket{a_j, b_j,p_m}{\prod_{j=1}^{g} \comm{a_j}{b_j} = \prod_{m=1}^{n} p_m}  .
\end{equation}

\subsection{Multiple covers}
\label{sec:multiple-covers}

An $N$--cover of a Riemann surface $\Sigma_{g,0}$ is a Riemann surface 
$\Sigma^N_{g,0}$.
By the Riemann--Hurwitz theorem,
its Euler characteristic is $\chi(\Sigma^N_{g,0}) = N \chi(\Sigma_{g,0})$
and its genus is \(g' = N \pqty{g - 1} + 1\).
The fundamental group of $\Sigma^N_{g,0}$ is an index $N$ subgroup of the
fundamental group of $\Sigma_{g,0}$, and the inequivalent covers of a
given surface are in one-to-one correspondence with the
inequivalent subgroups of $\pi_1(\Sigma_{g,0})$. 
In the following, we show how the subgroups can be classified in terms of maps \(\pi_1(\Sigma_{g,0}) \to S_N\) to the symmetric group of \(N\) elements.

\subsubsection{Free group of one element, $S^1$}

The relation between the symmetric group and the
covering maps  is easily understood in the case of $S^1$.
Its \(\pi_1\) is the free group with one element, $\langle a \rangle$.
Fix a point \(P\) on $S^1$ and consider the $N$--cover. This point   
is mapped to $N$ points $\set{P_1,\dots,P_N}$ on the covering space.
The action of the generator $a$ maps $P$ to itself, but in general it maps $(P_1,P_2,\dots,P_N)$ 
to a permutation $(P_{\sigma_a(1)},P_{\sigma_a(2)},\dots,P_{\sigma_a(N)})$.
The topology of the cover is fixed by the choice of the permutation $\sigma_a$.

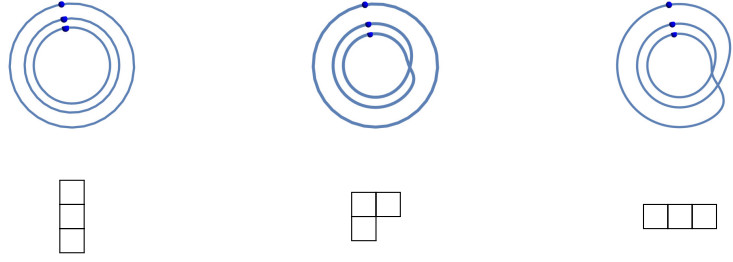
\begin{figure}
  \centering
  \begin{tikzpicture}
    \node at (-4,0) {\includegraphics[width=2cm]{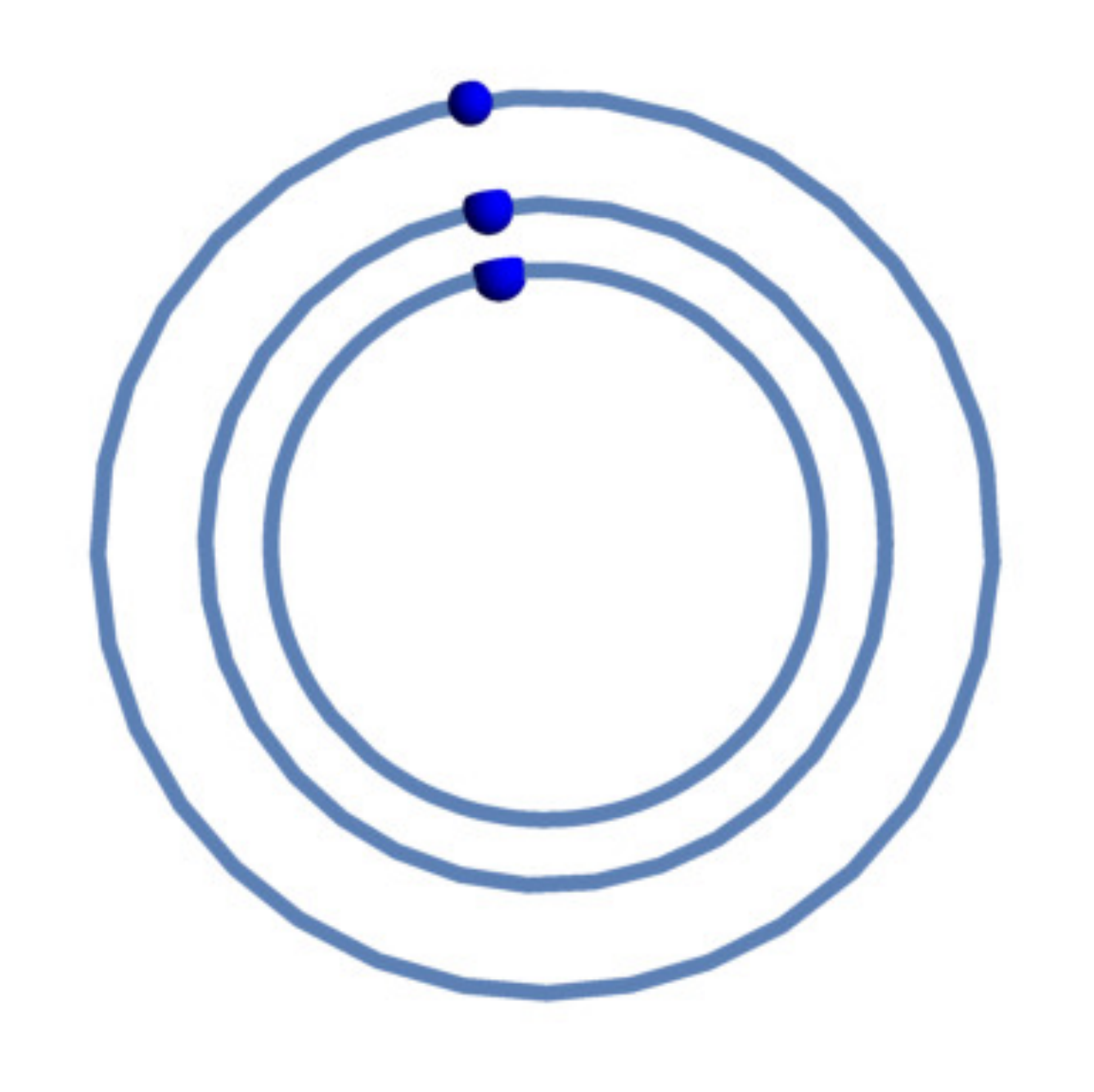}};
    \node at (0,0) {\includegraphics[width=2cm]{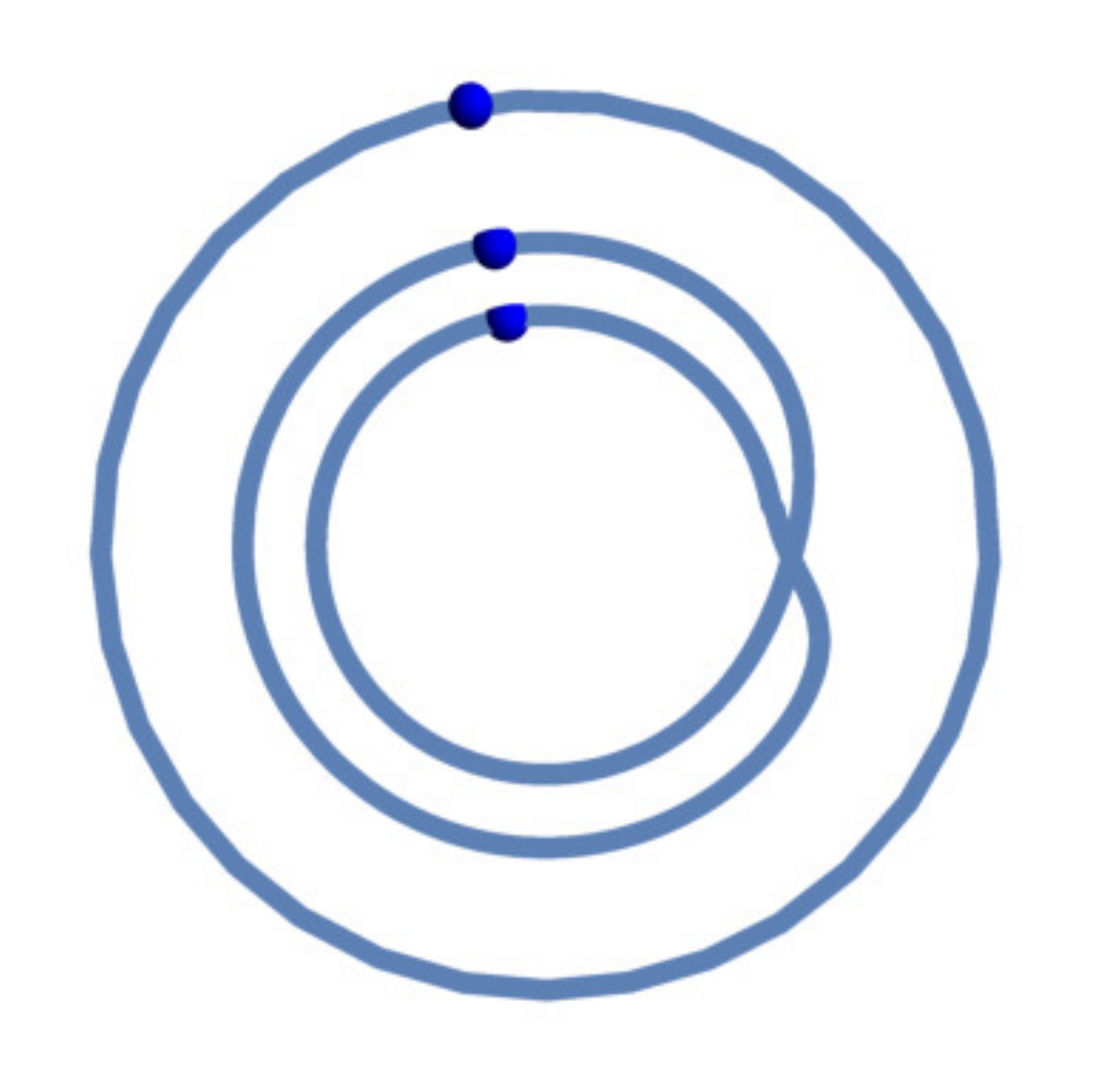}};
    \node at (4,0) {\includegraphics[width=2cm]{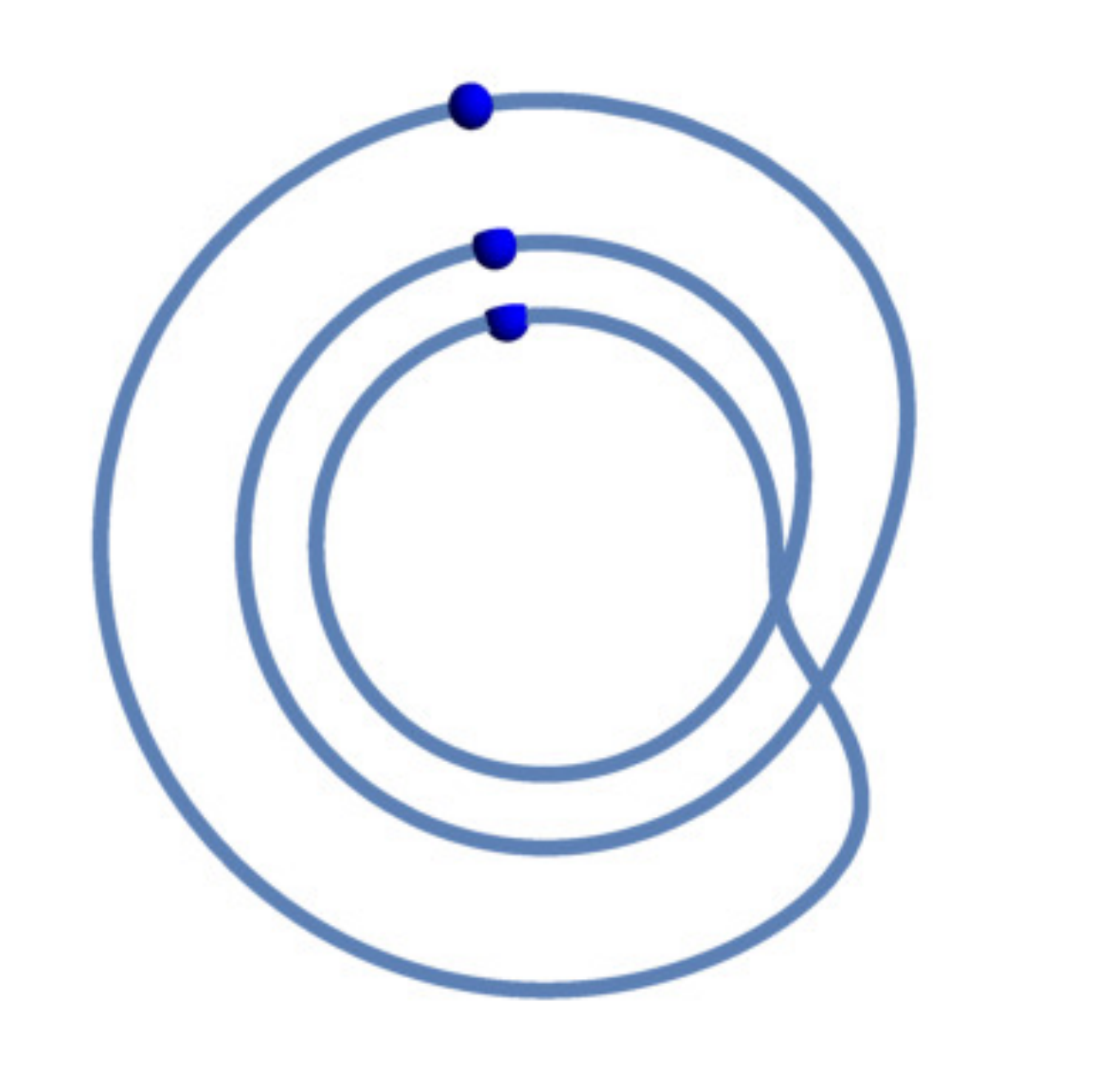}};

    \node at (-4,-2) {\ydiagram{1,1,1}};
    \node at (0,-2) {\ydiagram{2,1}}; 
    \node at (4,-2) {\ydiagram{3}};
  \end{tikzpicture}
  \caption{Triple covers of \(S^1\) and corresponding Young diagrams.
    Each line in the Young diagram maps to a connected component.}
  \label{fig:triple-S1}
\end{figure}

In Figure~\ref{fig:triple-S1}, the case \(N = 3\) is shown in detail. The image of a fixed point \(P\) in the base is given by three points \(\set{P_1, P_2, P_3}\) on the cover. Act with \(a\) on the base. The point \(P \) is mapped to itself, but in the three inequivalent covers, the three images are permuted as \(
\begin{psmallmatrix}
  P_1 & P_2 & P_3 \\ P_1 & P_2 & P_3 
\end{psmallmatrix}
\), \(
\begin{psmallmatrix}
  P_1 & P_2 & P_3 \\ P_2 & P_1 & P_3 
\end{psmallmatrix}
\) and \(
\begin{psmallmatrix}
  P_1 & P_2 & P_3 \\ P_3 & P_1 & P_2 
\end{psmallmatrix}
\). As usual, we can represent these three elements of \(S_3\) with the associated Young tableaux (respectively
\(\set{1,1,1}\), \(\set{2,1}\) and \(\set{3}\)). We have both connected and disconnected covers, and each row in the Young tableau corresponds to a connected component.%

\subsubsection{The torus}

As a second example, let us consider the two-torus $T^2$. 
This is a crucial example, because its geometry defines the lines in the $\mathcal{N}=4$ case as discussed in~\cite{Amariti:2015dxa} and many of the results obtained for this case are useful for more complicated orientable compact Riemann surfaces.
Like in the case above, multiple covers are classified by maps to the symmetric group and since 
there
are two generators, we have to introduce pairs of Young tableaux.

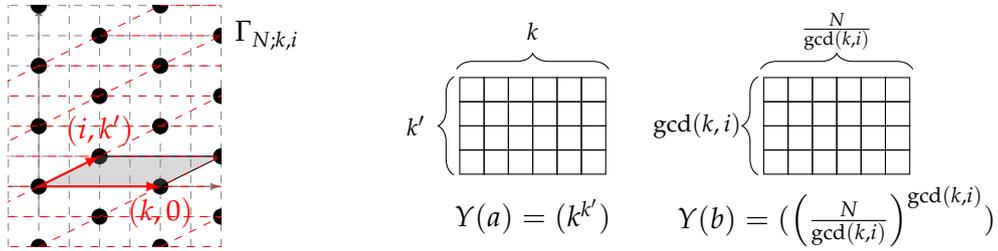
\begin{figure}
  \centering
  \begin{tikzpicture}
    \begin{scope}[shift={(-.5,0)}]
      \node at (0,0) {\ydiagram{6,6,6,6}};
      \draw [decorate,decoration={brace,amplitude=6pt},yshift=-.65cm](-1,1.4) -- (1,1.4) node[black,midway,yshift=0.5cm] {\footnotesize $k$};
      \draw [decorate,decoration={brace,amplitude=6pt},xshift=-1pt,yshift=-.65cm](-1,0) -- (-1,1.3) node[black,midway,xshift=-0.5cm] {\footnotesize $k'$};
      \node at (0,-1.2) {\(Y(a) = (k^{k'})\)};
    \end{scope}
    \begin{scope}[shift={(3.5,0)}]
      \node at (0,0) {\ydiagram{6,6,6,6}};
      \draw [decorate,decoration={brace,amplitude=6pt},yshift=-.65cm](-1,1.4) -- (1,1.4) node[black,midway,yshift=0.5cm] {\footnotesize $\frac{N}{\gcd(k,i)}$};
      \draw [decorate,decoration={brace,amplitude=6pt},xshift=-1pt,yshift=-.65cm](-1,0) -- (-1,1.3) node[black,midway,xshift=-.8cm] {\footnotesize ${\gcd(k,i)}$};
      \node at (0,-1.2) {\(Y(b) = (\pqty{\frac{N}{\gcd(k,i)}}^{\gcd(k,i)})\)};
    \end{scope}
    \begin{scope}[shift={(-6,0)},x=4cm, y=4cm]
      \foreach \name / \Xcoord / \Ycoord / \coeffk / \coeffkp / \coeffi in { 42/0/0/4/1/2} {
      \node (\name) at (\Xcoord, \Ycoord) {
        \begin{tikzpicture}[x=.4cm, y=.4cm]
          \lattice
        \end{tikzpicture}
      };
      \draw (\name) ++(.5,.3) node {\(\Gamma_{N;k,i}\)};
    }

    \end{scope}
  \end{tikzpicture}
  \caption{Descriptions of a connected double cover of \(T^2\), in terms of a lattice in \(\setZ^2\) and of a pair of permutations.}
  \label{fig:double-torus-connected-cover}
\end{figure}

A connected \(N\)--cover of the torus \(T^2_{N;k,i}\) is identified by a subgroup of \(\pi_1(T^2) = \setZ^2\). This is a lattice generated by the vectors \((k,0)\) and \((i, k')\) with the conditions \(k k' = N\) and \(0 \le i < k\). Equivalently,
\begin{equation}
  \pi_1(T^2_{N;k;i}) = \braket{a^k,  a^i b^{k'}}{ \comm{a^k}{b^{k'} a^i} = e} .
\end{equation}
This shows that the cover is again a torus, in accordance with the Riemann--Hurwitz theorem.
The corresponding map \(\pi_1(T^2) \to S_N\) is constructed as above. The origin in \(\setZ^2\) is 
mapped to $N$ integral points $\set{P_1,\dots,P_N}$ in the fundamental cell of $T^2_{N;k;i}$.
The action of the generator $a$ corresponds to a translation by one unit of the cell 
to the right. %
This defines a mapping $\sigma(a)$ of the set $\set{P_1,\dots,P_N}$ to itself.
By acting $k$ times with $a$, \emph{i.e.} by acting with $a^k$, the fundamental cell maps to itself.
This has a clear geometric meaning: all the cycles in the permutation $\sigma(a)$ have length $k$.
In terms of Young diagrams, $\sigma(a)$ is represented by a rectangle with $k$ columns and $N/k = k'$ rows.
The other generator is associated to the permutation $\sigma(b)$. This permutation
is obtained by finding the exponent in $b^p$ that maps the fundamental cell to itself.
In the sublattice, the origin is identified with all the 
points in the lattice of the form $m(k,0) + n(i,k')$, where $m$ and $n$ are integers and by definition \(b^p : (0,0) \mapsto (0,p)\).
This means that the length $p$ of the cycles in $\sigma(b)$ is given by the minimum integer $p$ such that
$m(k,0)+n(i,k')=(0,p)$ or equivalently $p=n k'$ and $mk+ni=0$.
The second equation gives $n/m = -k/i$, and $m$ and $n$ being both integer, we have $n = k/\gcd(k,i)$ and 
$m=i/\gcd(k,i)$. It follows that $p = N/\gcd(k,i)$, \emph{i.e.} the permutation $\sigma(b)$ corresponds to the conjugacy class represented by a box of $p=N/\gcd(k,i)$ columns and $\gcd(k,i)$ rows (see Figure~\ref{fig:double-torus-connected-cover}).
Observe that the number of columns in one diagram is greater or equal to the number of rows in the other.
\input{lattice-picture}

\bigskip

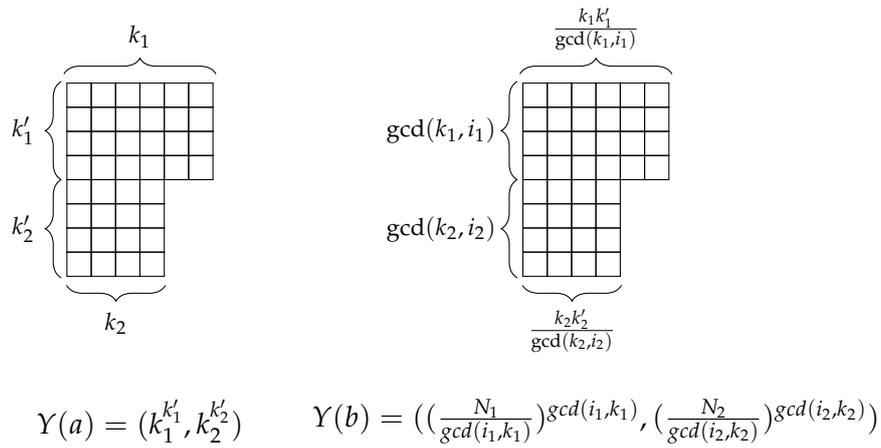
\begin{figure}
  \centering
  \begin{tikzpicture}
    \begin{scope}[shift={(-3,0)}]
      \node at (0,0) {\ydiagram{6,6,6,6,4,4,4,4}};
      \draw [decorate,decoration={brace,amplitude=6pt}](-1,1.4) -- (1,1.4) node[black,midway,yshift=0.5cm] {\footnotesize $k_1$};
      \draw [decorate,decoration={brace,amplitude=6pt},xshift=1pt](.3,-1.4) -- (-1,-1.4) node[black,midway,yshift=-0.5cm] {\footnotesize $k_2$};
      \draw [decorate,decoration={brace,amplitude=6pt},xshift=-1pt](-1,0) -- (-1,1.3) node[black,midway,xshift=-0.5cm] {\footnotesize $k_1'$};
      \draw [decorate,decoration={brace,amplitude=6pt},xshift=-1pt](-1,-1.3) -- (-1,0) node[black,midway,xshift=-0.5cm] {\footnotesize $k_2'$ };
      \node at (0,-3.2) {\(Y(a) = (k_1^{k_1'},k_2^{k_2'})\)};
    \end{scope}
    \begin{scope}[shift={(3,0)}]
      \node at (0,0) {\ydiagram{6,6,6,6,4,4,4,4}};
      \draw [decorate,decoration={brace,amplitude=6pt}](-1,1.4) -- (1,1.4) node[black,midway,yshift=0.6cm] {\footnotesize $\frac{k_1 k_1'}{\gcd(k_1,i_1)}$};
      \draw [decorate,decoration={brace,amplitude=6pt},xshift=1pt](.3,-1.4) -- (-1,-1.4) node[black,midway,yshift=-0.6cm] {\footnotesize $\frac{k_2 k_2'}{\gcd(k_2,i_2)}$};
      \draw [decorate,decoration={brace,amplitude=6pt},xshift=-1pt](-1,0) -- (-1,1.3) node[black,midway,xshift=-1cm] {\footnotesize ${\gcd(k_1,i_1)}$};
      \draw [decorate,decoration={brace,amplitude=6pt},xshift=-1pt](-1,-1.3) -- (-1,0) node[black,midway,xshift=-1cm] {\footnotesize ${\gcd(k_2,i_2)}$};
      \node at (0,-3.2) {$Y(b) = ((\frac{N_1}{gcd(i_1,k_1)})^{gcd(i_1,k_1)},(\frac{N_2}{gcd(i_2,k_2)})^{gcd(i_2,k_2)})$};
    \end{scope}
  \end{tikzpicture}
  \caption{Pair of permutations describing a double cover of \(T^2\) with two connected components.}
  \label{fig:double-torus-disconnected-cover}
\end{figure}

A \emph{disconnected cover} of the torus is represented by the union of rectangular Young diagrams.
For example, if the cover has two connected components they are both  tori, $T_{N_1;k_1;i_1}^2$ 
and  $T_{N_1;k_1;i_1}^2$, such that $N_1 + N_2 = k_1 k_1' + k_2 k_2' = N$ and the Young diagrams are  $Y(a) = (k_1^{k_1'},k_2^{k_2'})$
and $Y(b) = ((\frac{N_1}{gcd(i_1,k_1)})^{gcd(i_1,k_1)},(\frac{N_2}{gcd(i_2,k_2)})^{gcd(i_2,k_2)})$ 
respectively (see Figure~\ref{fig:double-torus-disconnected-cover}).

\FloatBarrier

\printbibliography

\end{document}

%% file: lattice-picture.tex
\newcommand*{\lattice}{
    \coordinate (Origin)   at (0,0);
    \coordinate (XAxisMin) at (-1,0);
    \coordinate (XAxisMax) at (6,0);
    \coordinate (YAxisMin) at (0,-1);
    \coordinate (YAxisMax) at (0,6);
    \draw [thin, gray,-latex] (XAxisMin) -- (XAxisMax);%
    \draw [thin, gray,-latex] (YAxisMin) -- (YAxisMax);%

    \clip (-1,-2) rectangle (6,6); %

    \draw[style=help lines,dashed] (-2,-2) grid[step=1] (7,7);

    \pgftransformcm{\coeffk}{0}{\coeffi}{\coeffkp}{\pgfpoint{0}{0}}

    \coordinate (Xone) at (1,0);
    \coordinate (Xtwo) at (0,1);

    \foreach \x in {-7,-6,...,7}{%
      \foreach \y in {-7,-6,...,7}{%
        \node[draw,circle,inner sep=2pt,fill] at (1*\x,1*\y) {};
      }
    }
    \filldraw[fill=gray, fill opacity=0.3, draw=black] (Origin)
        rectangle ($(Xone)+(Xtwo)$);

    \draw[style=help lines,dashed, red] (-2,-2) grid[step=1] (7,7);

    \draw [thick,-latex,red] (Origin) -- (Xone) node [below] {$(k,0)$};
    \draw [thick,-latex,red] (Origin) -- (Xtwo) node [above] {$(i,k')$};
}